\documentclass[epj,final]{svjour}
\usepackage{ulem}
\usepackage{color}
\usepackage{graphicx,url}
\usepackage{bm} 
\usepackage{amssymb}

\begin{document}
\title{Line shape analysis of the K${\bm\beta}$ transition in muonic hydrogen}
%
\normalem   
\author{D.\,S.\,Covita\inst{1,2}\thanks{\emph{present address:} Bosch Termotecnologia, S.A., EN 16 - Km 3,7, 3800-533 Cacia Aveiro, Portugal}, 
        D.\,F.\,Anagnostopoulos\inst{3},
        H.\,Fuhrmann\inst{4}, 
        H.\,Gorke\inst{5}, 
        D.\,Gotta\inst{6}\thanks{\emph{corresponding author:} d.gotta@fz-juelich.de}, 
        A.\,Gruber\inst{4}\thanks{\emph{present address:}
        Dept. of Nuclear Medicine, Vienna General Hospital, A-1090 Vienna, Austria}, 
        A.\,Hirtl\inst{4}\thanks{\emph{present address:} Atominstitut, TU Wien, A-1020 Vienna, Austria}, 
        T.\,Ishiwatari\inst{4},
        P.\,Indelicato\inst{7,8,9},
        T.\,S.\,Jensen\inst{2,10}\thanks{\emph{present address:} 
        Ringkj$\o$bing Gymnasium, Vasevej 24, 6950 Ringkj$\o$bing,  Denmark},
        E.-O.\,Le Bigot\inst{7},
        V.\,E.\,Markushin\inst{2}\thanks{\emph{present address:} 
        Department IT, Paul Scherrer Institut, CH-5232 Villigen, Switzerland},
        M.\,Nekipelov\inst{6}\thanks{\emph{present address:} 
        WTI GmbH, D-52428 J\"ulich, Germany},
        V.\,N.\,Pomerantsev\inst{11},
        V.\,P.\,Popov\inst{11},
        J.\,M.\,F.\,dos Santos\inst{1},
        Ph.\,Schmid\inst{4},
        L.\,M.\,Simons\inst{2},
        M.\,Theisen\inst{6}\thanks{\emph{present address:} 
        Kayoom GmbH, D-52353 D\"uren, Germany},
        M.\,Trassinelli\inst{12},
        J.\,F.\,C.\,A.\,Veloso\inst{13},
   \and J.\,Zmeskal\inst{4}
}                     
\institute{     LIBPhys, Physics Department, University of Coimbra, P-3004-526 Coimbra, Portugal
           \and Laboratory for Particle Physics, Paul Scherrer Institut, CH-5232 Villigen, Switzerland
           \and Department of Materials Science and Engineering, University of Ioannina, GR-45110 Ioannina, Greece 
           \and Stefan Meyer Institute for Subatomic Physics, Austrian Academy of Sciences, A-1090 Vienna, Austria
           \and Zentralinstitut f\"{u}r Elektronik, Forschungszentrum J\"{u}lich GmbH, D-52425 J\"{u}lich, Germany
           \and Institut f\"ur Kernphysik, Forschungszentrum J\"ulich, D-52425 J\"ulich, Germany
           \and Laboratoire Kastler Brossel,  Sorbonne Universit\'es, UPMC Univ. Paris 06, Case 74; 4, place Jussieu, 75005 Paris, France
\and
Laboratoire Kastler Brossel,  CNRS, 75005, Paris, France
\and
Laboratoire Kastler Brossel, D\'epartement de Physique de l'\'Ecole Normale Sup\'erieure,  24 Rue Lhomond, 75005, Paris, France 
           \and Institut f\"ur Theoretische Physik Universit\"at Z\"urich, CH-8057 Z\"urich, Switzerland
           \and Skobeltsyn Institut of Nuclear Physics, Lomonosov Moscow State University, 119991 Moscow, Russia
           \and Institut des NanoSciences de Paris, CNRS-UMR 7588, Sorbonne Universit\'es, UPMC Univ Paris 06, 75005, Paris, France
           \and I3N, Department of Physics, Aveiro University, P-3810 Aveiro, Portugal}
\date{Received: date / Revised version: date}
%
\abstract{
The K$\beta$ transition in muonic hydrogen was measured with a high-resolution crystal spectrometer. The spectrum is shown to be sensitive to the ground-state hyperfine splitting, the corresponding triplet-to-singlet ratio, and the kinetic energy distribution in the $3p$ state.  The hyperfine splitting and triplet-to-singlet ratio are found to be consistent with the values expected from theoretical and experimental investigations and, therefore, were fixed accordingly in order to reduce the uncertainties in the further reconstruction of the kinetic energy distribution. The presence of high-energetic components was established and quantified in both a phenomenological, {\it i.\,e.} cascade-model-free fit, and in a direct deconvolution of the Doppler broadening based on the Bayesian method.} 
\PACS{{36.10.-k, 07.85.Nc}{Exotic atoms, X-ray spectrometers}} 
%
\authorrunning{D.\,S.\,Covita {\it et al.}}
\titlerunning{Line shape analysis of the K$\beta$ transition in muonic hydrogen}
\maketitle
\section{Introduction}\label{sec:intro}

Exotic atoms are formed when heavy negatively charged particles like muons, pions, kaons, or antiprotons are slowed down to the eV range and captured in the Coulomb field of the nucleus. They provide the opportunity to study QED, weak, or strong interactions but also collision-induced processes at quite distinctive conditions, because the distance scale in the case of exotic hydrogen is smaller by a factor $m_{e}/m_{X}$ while the energy scale is larger by a factor $m_{X}/m_{e}$ in comparison with usual atoms. Here, $m_{e}$ and $m_{X}$ denote the mass of electron and captured particle $X$, respectively. 

An exotic atom formed with hydrogen isotopes is exceptional, because it is electrically neutral. Hence, during its life time it is able to penetrate deeply into the atomic shells when colliding with other molecules. Various collisional processes strongly influence the development of the atomic de-excitation cascade, which manifests experimentally in a strong density dependence of the K X-ray yields\cite{And84,Bre98,Lau98,Bor80}. Amongst others, in the Coulomb de-excitation process the  energy released is transferred to kinetic energy\,\cite{Bra78} leading to a Doppler broadening for subsequently emitted X-rays. 

First evidence for such high-velocity exotic atoms was found in the charge-exchange reaction $\pi^{-}p\to\pi^0 n$ with stopped pions as Doppler broadening of the time-of-flight  distribution of the monoenergetic neutrons\,\cite{Czi63,Bad01}. For X-ray transitions of exotic atoms, a direct observation of Doppler broadening became feasible by using ultimate-resolution X-ray spectroscopy\,\cite{Sie00,Got04,Cov09}.

The measurement of the K$\beta$ X-ray transition in muonic hydrogen described here was embedded in a series of measurements on pionic hydrogen and deuterium\,\cite{PSI98,Got08,Str10,Str11}. These experiments aimed at a new high-precision determination of the hadronic line shift and broadening of K transitions in order to determine the pion-nucleon and pion-deuterium scattering lengths\,\cite{Gas08}. However, the precise determination of the level widths in pionic hydrogen is hindered by a significant Doppler broadening of X-ray lines stemming from Coulomb de-excitation. 

Muonic hydrogen undergoes ---apart from hadronic\linebreak effects--- similar processes in the atomic cascade. Hence,  a quantitative study of the Doppler-induced broadening reveals details of the Coulomb de-excitation to be compared with recent developments in the theoretical description of the kinetics of the atomic cascade\,\cite{Jen02c,JPP07,Pop11,Pop12,LLM16} without the additional difficulty of unfolding the hadronic broadening. The knowledge gained can then eventually be applied to the pionic hydrogen case.

Compared to earlier publications resulting from the present series of experiments \cite{Got08,Str10,Str11,Got08a,Str09,Got12,Hen14}, the evaluation process is extended considerably. In addition to the so-called frequentist approach used before, the Bayesian method was used in this work as well. This was mainly motivated by the analysis of pionic deuterium  where a $\chi^2$ analysis was used in order to achieve the maximum probability that the data match the chosen model\,\cite{Str10,Str11}. 

The $\chi^2$ method may fail especially because a bias influencing the extracted parameter values may occur, which stems from the ---principally unknown--- deviation of the probability distribution owing to the data itself and the one assumed for the model. A possible bias is then determined by applying the same model  as used for the analysis to a series of Monte-Carlo simulations and by comparing input and output parameters.

Such a lengthy procedure may be circumvented in using Bayesian methods which are claimed to be free from bias effects in parameter estimation\,\cite{Cou95}. They may suffer from different  drawbacks, however, {\it e.\,g.} the influence of the choice of a prior distribution. 

Both methods were exploited in the determination of parameters such as the energy splitting $\mathrm{\Delta}E_{\mathrm{hfs}}$ and the ratio $R_{\mathrm{T/S}}$ of the population of the hyperfine levels of the ground state. By 
comparing the results of an earlier publication \cite{Cov09,Cov09a} with the  Bayesian approach, a consistency check is established. It goes as well in line with a recent assessment of both methods recommending to use  both approaches as  good practice  in evaluating data\,\cite{Bab12}. A comparison of the features of the two evaluation methods is given in apps.\,A.1 and A.2. 

The paper outlines the processes occurring during the de-excitation cascade and its theoretical treatment\linebreak
 in sec.\,\ref{sec:cascade}. The quality of the results decisively depends on an almost background-free and simultaneous measurement of the energy interval covering the complete line shape as well as a very precise knowledge of the spectrometer response. Therefore, the experimental set-up is described in quite detail in sec.\,\ref{sec:exp}. The procedure of extraction of the energy spectrum from the raw data is described in sec.\,\ref{sec:analysis}. The results of frequentist\,\cite{Cov09} and Bayesian method are compared  in sec.\,\ref{sec:results}.
 
The result of a deconvolution procedure, which aims at the unfolding of the kinetic energy distribution directly from the measured spectrum, is presented in sec.\,\ref{subsec:deconv}. The principle of the Bayesian deconvolution is outlined in app.\,B, and its robustness is demonstrated with two examples for the kinetic energy distribution including one for an advanced cascade model.


\section{Atomic cascade}\label{sec:cascade}

\begin{figure}[h]
\begin{center}
\resizebox{0.48\textwidth}{!}{\includegraphics{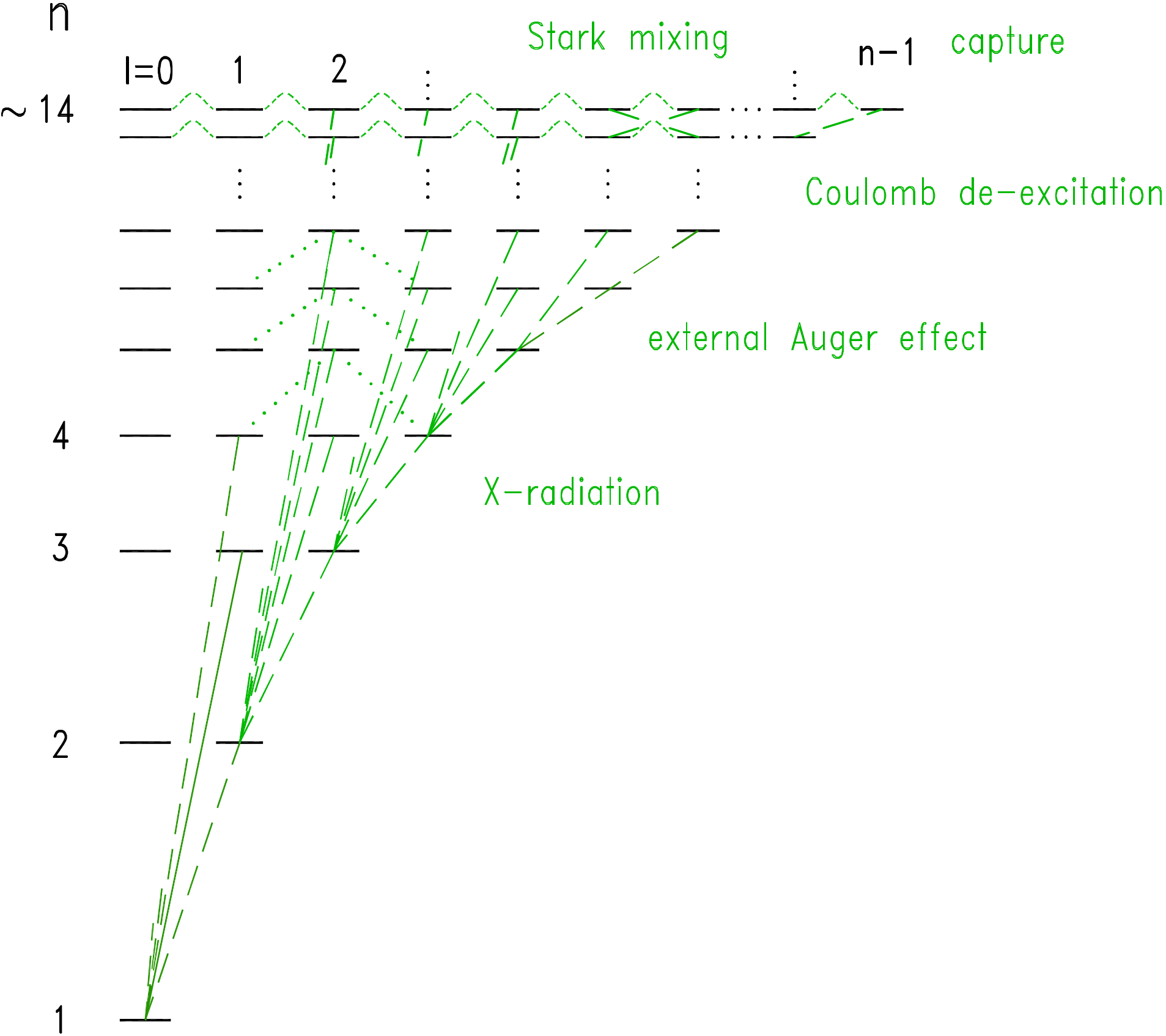}}
\caption{De-excitation cascade in muonic hydrogen. The spin-averaged energy of the $\mu$H K$\beta$ X-ray transition $(3p-1s)$ is calculated to be (2249.461\,$\pm$\,0.001)\,eV.}
\label{figure:fig1_muH_cascade}
\end{center}
\end{figure}

Formation of exotic hydrogen occurs in highly excited atomic states with a wide spread of the  principle quantum number peaking at about $n_c=\sqrt{\mu_X/m_e}$, where $\mu_X$ is the $X^-p$ reduced mass\,\cite{Har90}. In the case of muons, $n_c\approx 14$ (Fig.\,\ref{figure:fig1_muH_cascade}). The initial distributions for $n$ and  angular momentum $\ell$ have been calculated including modifications arising from capture by hydrogen molecules\,\cite{Kor89,Kor92,Coh99,Coh04}.

The further evolution of muonic hydrogen is determined by the competition of radiative transitions and  collisional processes.  Collisional processes are external Auger effect, elastic and inelastic scattering, Stark transitions,  dissociation of H$_2$ molecules, and Coulomb de-excitation. The cascade time is estimated from cascade calculations to be of the order of 0.1\,ns for densities around $\varphi = 10^{-2}$ (in units of the liquid hydrogen density = $4.25\cdot 10^{22} {\rm atoms/cm}^3$)\,\cite{Bor80}, $\it i.\,e.$ much shorter than the muon's lifetime. Consequently, almost all muons reach the atomic ground state $1s$.

As earlier experimental X-ray studies were restricted to the measurement of line yields and  reaction products owing to the weak interaction or muon-catalysed fusion, the Doppler broadening detectable in present-day experiments yields additional information on the kinetics  developing during the atomic cascade\,\cite{Bad01,Sie00,Got04,Cov09,Got08,Kot01,Poh10,Poh01,Sch01,Poh06}. Therefore, a realistic theoretical description of such a dynamics along with the experimental progress became essential. 

\subsection{Developments in the theoretical approach}\label{subsec:cascade_theory}

The first detailed theoretical study of the atomic cascade in exotic hydrogen-like  ($\pi^{-}p$ and $K^{-}p$) atoms was  performed by Leon and Bethe \cite{Leo62} more than fifty years ago. In this paper and later in more refined models \cite{Bor80,Jen03,Ter97}, the rates of the collisional processes Stark mixing and external Auger effect  were calculated, respectively, in the semi-classical and Born approximation {\it at fixed kinetic energy} ($\sim 1$ eV) to simulate the atomic cascade. In these calculations,  scaling factors for Stark mixing and external Auger effect were introduced and kinetic energy was treated as a fit parameter  and, indeed, often found to be of the order of 1\,eV. 

However, the energy distributions of $\mu^{-}p$, $\mu^{-}d$ and $\pi^{-}p$ atoms, measured by applying different time-of-flight methods \cite{Bad01,Cra91,Abb97,Poh01a}, revealed that their kinetic energy\linebreak changes dramatically during the cascade. In particular, the existence of high-energetic components of up to 200\,eV was established by means of the time-of-flight of neutrons stemming from the reaction $\pi^{-}p\rightarrow \pi^{0}n$ (see \cite{Bad01}) and muonic hydrogen diffusion experiments \cite{Abb97,Poh01a}. In the case of $\mu^{-}p$, the kinetic energies corresponding to the low-lying $\mathrm{\Delta}n=1$ Coulomb de-excitation transitions $(7-6)$, $(6-5)$, $(5-4)$, $(4-3)$, and $(3-2)$ are 9, 15, 27, 58, and 166\,eV, respectively.  

Consequently, the evolution of both kinetic energy and $(n,\ell)$ distributions should be taken into account during the whole atomic cascade. Hence, the \textit{extended standard cascade model} (\textit{ESCM}) \cite{Jen02c,Mar94} was introduced, which allows to trace the kinetic energy through the complete de-excitation cascade.

In a further step, analysis and interpretation of such observations require a more sophisticated approach based on the reliable and self-consistent calculation of all collision-induced processes involved in the de-excitation cascade and should be described in a fully quantum-mechanical approach covering a wide range of both quantum numbers and collision  energies.

Significant improvements over previous cascade calculations\,\cite{Jen02c,Jen03} were achieved during last decade\,(see \cite{JPP07,Pop11,Pop12,LLM16} and references therein). The present version of the \textit{ESCM} includes: new theoretical results for the collisional processes, initial distributions of muonic atoms in quantum numbers and the laboratory kinetic energy at the instant of their formation, and the thermal motion of the target. In particular, the comprehensive sets of differential and integrated cross sections of collision-induced processes involved in the atomic cascade were calculated within a fully quantum-mechanical approach in the wide range of quantum numbers and kinetic energy of muonic atoms that are of interest to reliable cascade studies.

\subsection{Cascade processes in muonic hydrogen}
 
Only at very low density, the de-excitation cascade is dominated by the radiative transitions (see, {\it e.g.}  \cite{Pop11}). Radiative de-excitation is sufficiently described by the electric-dipole transitions, where the initial and final values of the internal orbital angular-momentum satisfy the selection rule $\mathrm{\Delta}\ell = \left|\ell_{i}-\ell_{f}\right|=1$. Since the recoil energies are of the order of a few meV, the radiative de-excitation hardly affects the kinetic energy distribution of exotic hydrogen.
 
The main collisional processes are external Auger effect, elastic scattering, Stark transitions, and Coulomb de-excitation. Already at modest densities (above $\varphi\approx 10^{-5}$) the collisional processes with rates proportional to the target density determine the development of the cascade. External Auger effect does not change significantly the kinetic energy of the exotic atom since the transition energy is carried away mainly by the electron. The Auger transitions with the minimal $\mathrm{\Delta}n = n_{i}-n_{f}$ needed for ionization are the most probable with rates reaching their maximum for $\mathrm{\Delta}n = 1$ and rapidly decreasing both for decreasing $n$ and for $\mathrm{\Delta}n\,>\,1$ transitions. 

In collisional processes
 \begin{equation} \label{eq1}
(X^- p)_{n\ell} + H_{1s}\rightarrow(X^- p)_{n'\ell'} +H_{1s},
\end{equation}
elastic scattering ($n'=n,\,\ell'=\ell$), Stark transitions ($n'=n,\,\ell'\neq \ell$), and Coulomb de-excitation ($n'\!< \!n$) essentially change the kinetic energy of the exotic atom. Coulomb de-excitation plays the most important role because in this case, an acceleration of the exotic atom occurs due to the fact that the energy of the transition $n\to n'$ is converted into kinetic energy and shared between the colliding objects (exotic and target atoms).

Coulomb de-excitation has been studied in the two-state semi-classical model (usually for $n \gg 3$)~\cite{Bra78}, within the advanced adiabatic~\cite{Kra01}, or the classical-trajectory\linebreak Monte-Carlo\,\cite{Jen02b} approaches. However, for low-lying $n$ states and low collision energies, the approaches described in\,\cite{Bra78,Jen02b} cannot be expected to give a reliable description of the Coulomb de-excitation. The rates obtained in adiabatic approximation~\cite{Kra01} for the $n\rightarrow n-1$ transitions ($n=3-5$) are more or about one order of magnitude smaller than the ones given in\,\cite{Bra78} and, consequently, too small to explain the observed kinetic energy distribution\,\cite{Bad01} of $(\pi^- p$) atoms at the instant of the charge exchange reaction $\pi^- p \to \pi^0 n$. This reaction occurs predominantly in the range $n=3-8$. 

In order to improve the cascade description, the dynamics of the collision of excited exotic systems with hydrogen 
atoms has been studied in the framework of the close-coupling approach where elastic scattering, Stark and Coulomb transitions can be treated in a unified manner\,\cite{Pop11,Kor05,Pom06a,Pom06b,Pop08}. Differential and integral cross sections were calculated in a wide range of the principle quantum number and relative kinetic energies for the case of muons, pions, and antiprotons by taking into account vacuum polarization and, for hadronic atoms, strong interaction shifts and broadenings. It was shown in \cite{Kor05,Pom06a,Pom06b,Pop08} that the $\mathrm{\Delta} n= 1$ transitions dominate the Coulomb de-excitation, but transitions with $\mathrm{\Delta}n>1$ contribute substantially $(20\%-40\%)$ to the total Coulomb de-excitation cross sections for $n\geq 4$ at all 
the energies under consideration.

Stark transitions, preserving the principal quantum number $n$, affect the population of the ($n,\ell$) subshells. Together with the elastic scattering they can decelerate the exotic atoms thus influencing their kinetic energy distribution during the cascade. Starting from the classical paper by Leon and Bethe\,\cite{Leo62}, Stark transitions were often treated in the semi-classical straight-line-trajectory approximation (see~\cite{Ter97} and references therein). The first fully quantum-mechanical study of the elastic scattering and Stark mixing cross sections was performed within the fully quantum-mechanical adiabatic approach\,\cite{Pop9699,Gus99}. Later, the cross sections were also calculated in the framework of the close-coupling model and various versions of the semi-classical approximation \cite{Jen02a}, where the Coulomb interaction of the exotic atom with the hydrogen-atom field is modelled by the screening dipole approximation. This approximation is not justified at low collision energies below some value, which depends on the principal quantum number $n$ (see \cite{Gus99}). Thus, the latter approach as well as various modifications of the semiclassical model\,\cite{Bor80,Ter97,Jen02a} can result in uncontrolled uncertainties in the low-energy region where only a few partial waves are important.

\begin{figure}[h]
\begin{center}
\resizebox{0.5\textwidth}{!}{\includegraphics{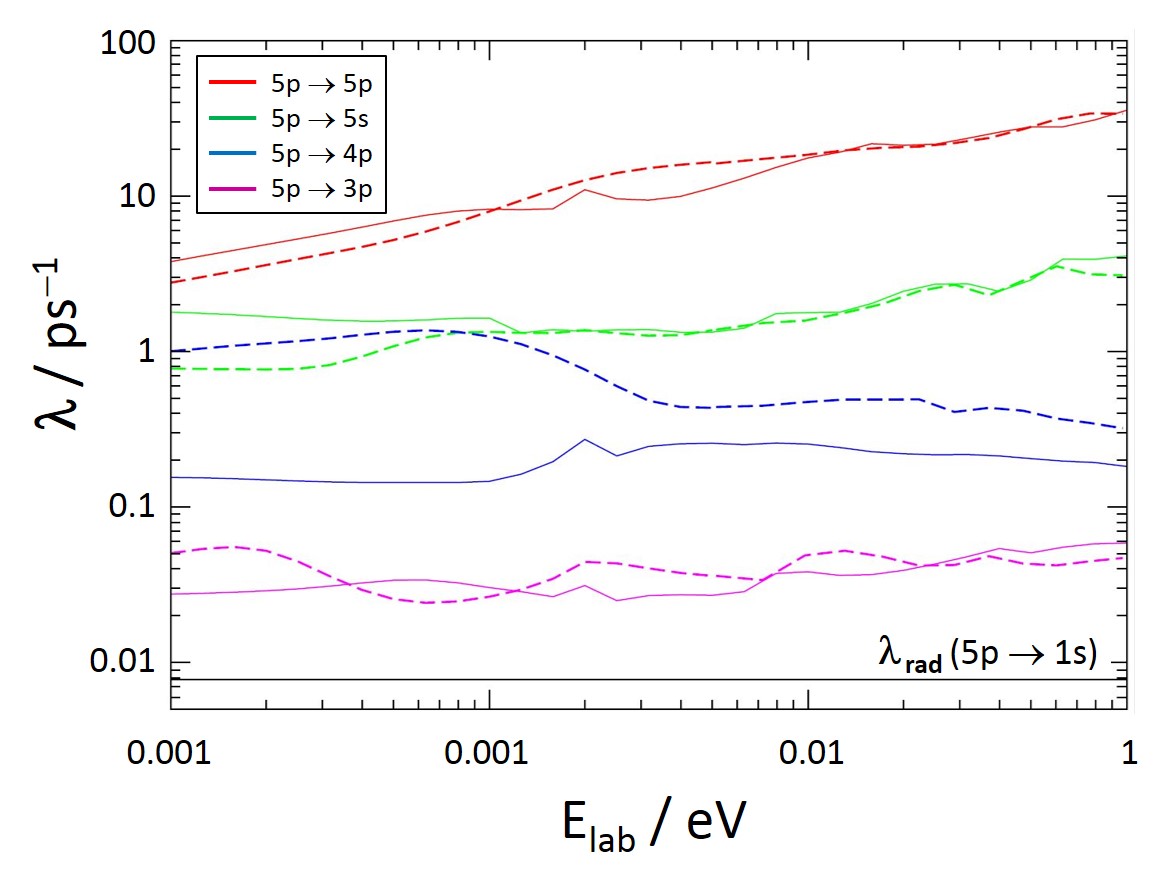}}
\caption{Energy dependence of the collision rates for the $5p$ state of muonic hydrogen calculated at liquid hydrogen density for two variants of the basis set: $n_{\mathrm{max}}=5$ (solid) and $n_{\mathrm{max}}=20$ (dashed). Rates of collisional processes scale practically linearly with density. The radiative transition rate $\lambda_{5p\rightarrow 1s}$ is shown for comparison. 
}
\label{figure:fig2_muH_n5_rates}
\end{center}
\end{figure}

A typical example of the collisional rates calculated in the fully quantum-mechanical close-coupling approach is shown in fig.\,\ref{figure:fig2_muH_n5_rates} for muonic hydrogen in the states with $n =5$ at hydrogen density $\varphi=1$ and and kinetic energies $E_{\mathrm{lab}}=0.001 - 1$~eV  for two variants of the basis states: $n_{\mathrm{max}}=5$ and $n_{\mathrm{max}}=20$ (for details and discussion see\,\cite{Pop11,LLM16}).
 
The extended basis set ($n_{\mathrm{max}}=20$) includes in addition to the open channels all exotic-atom states with principal quantum numbers from $n=6$ up to $n=20$ corresponding to the closed channels of the scattering problem. Obviously, the effect of closed channels is very important and significantly changes the collision rates especially at low energies. In particular, it results in an increase of more than a factor of two for the rate of the ($5-4$) and of about 30\% for the rate of the ($5-3$) Coulomb transition.

\subsection{The improved extended standard cascade model}

\textit{ESCM}\,\cite{Jen02c,Mar94} introduced the evolution of the kinetic energy during de-excitation cascade as well as the scattering from molecular hydrogen at high $n$ beyond a purely phenomenological approach. However, the first \textit{ESCM} predictions based on older cross section results were not able to provide a precise description of the experimental data from kinetic-energy distributions at the time of $X$-ray emission from a specific level in muonic and pionic atoms or at the time of neutron emission in the charge-exchange reaction (neutron time-of-flight experiment). 

Including the results obtained from self-consistent treatment of elastic scattering, Stark transitions, and Coulomb de-excitation in a framework of the close-coupling approach led to a substantial improvement\,\cite{JPP07,Pop11,Pop12,LLM16}. In such quantum mechanical calculations of differential and integrated cross sections for $n\ell\rightarrow n'\ell'$ collisional transitions, an accurate description is obtained of the kinetics of the atomic cascade without employing any fitting parameters.

For the calculation, the atomic cascade (as in the previous \textit{ESCM}, see {\it e.\,g.} \cite{Jen02c} and refs. therein), is divided into the classical and the quantum mechanical domain for high and low $n$ states, respectively. In the upper part of the cascade ($n \geq 9$), the classical-trajectory Monte-Carlo results were included (see for details \cite{Jen02c,Pop07}) with the molecular structure of the target taken into account for elastic scattering, Stark mixing, and Coulomb de-excitation. The resulting distributions in quantum numbers and kinetic energy were used as input for the next stage of the atomic cascade. In the quantum mechanical domain ($n\leq 9$), the differential and integrated  cross sections for  all $n\ell\rightarrow n'\ell'$ transitions (besides Auger and radiative de-excitation) were calculated in the fully quantum-mechanical close-coupling approach (see \cite{Pop11,LLM16}) for the initial states with $n\leq 8$ and relative energies $E\geq 0.001$~eV. 

The initial distributions in the quantum numbers $n$ and $\ell$ calculated in \cite{Kor89,Kor92,Coh99,Coh04} take into account the molecular effects of the target and predict an initial $n_i$ distribution with the maximum shifted towards
lower $n\cong 11 < n_c$ values. Besides, the populations of   ($n,\ell$) sub-levels have a non-statistical distribution over $\ell$ for each value of the principal quantum number. In the present cascade calculations, the effects of the initial $(n,\ell,E)$ distributions and target temperature are properly taken into account as given in\,\cite{Pop11}. In order to obtain good statistics, the cascade calculations were performed with $10^7$ events.

\subsection{K X-ray yields in muonic hydrogen}
   
\begin{figure}[h]
\begin{center}
\resizebox{0.48\textwidth}{!}{\includegraphics{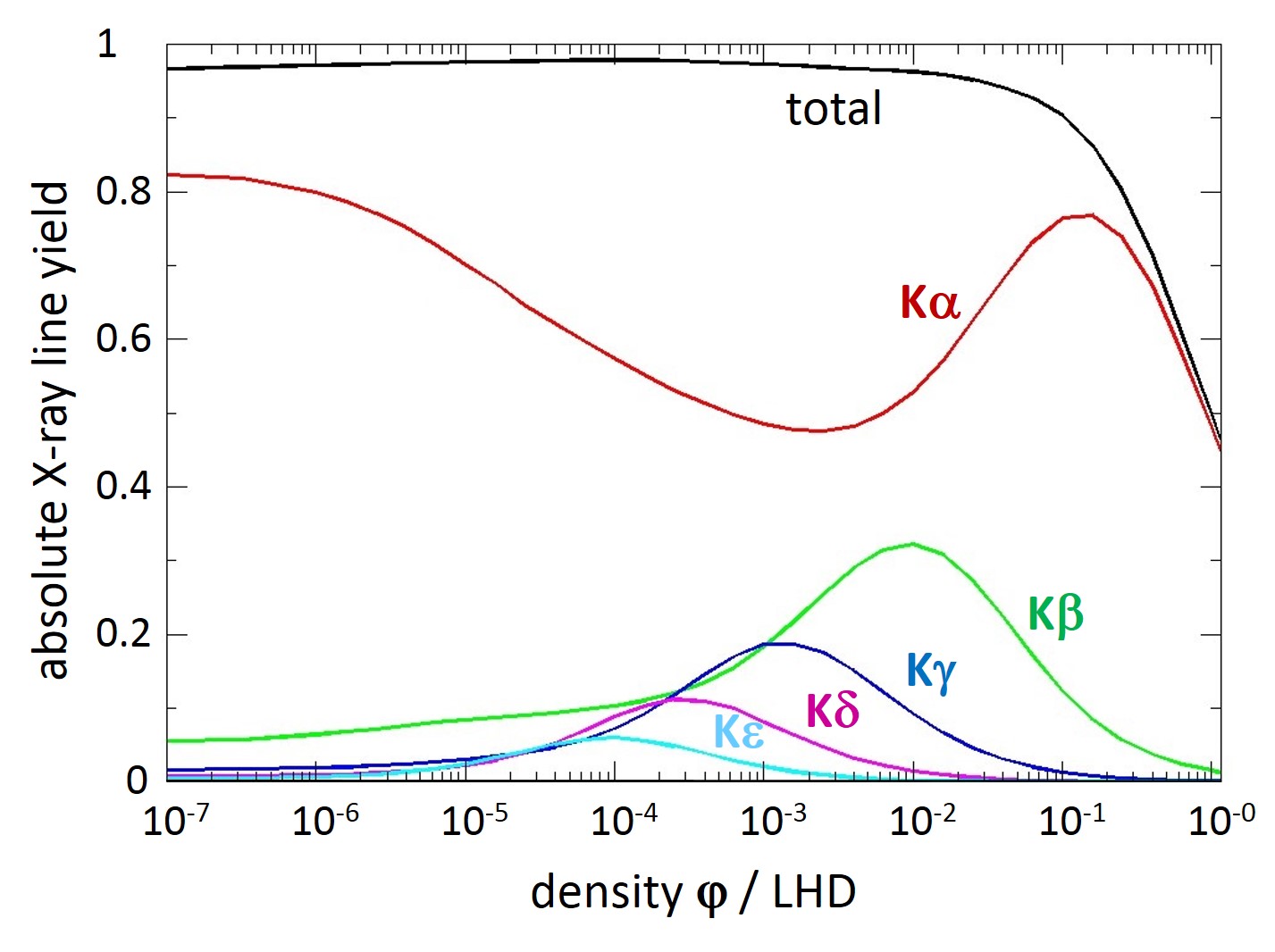}}
\caption{Calculated density dependence of line yields of $K$ transitions in muonic hydrogen at a target temperature $T=30$\,K.}
\label{figure:fig3_K_yield_absolute}
\end{center}
\end{figure}

The total K X-ray yield reaches 96\% and is approximately constant from lowest densities up to $\varphi = 10^{-2}$. For liquid hydrogen, it drops to about 50\%\,\cite{LLM16} because of the rapidly increasing non-radiative de-excitation (fig.\,\ref{figure:fig3_K_yield_absolute}).

In contrast to the total K X-ray yield, the yields of individual K lines $Y_{Ki}$ ($i = \alpha, \beta$, ...) exhibit a complicated density dependence due to the competition between the various collisional and non-collisional processes during the atomic cascade. At low target densities, the atomic cascade is dominated by radiative de-excitation proceeding through circular states $(n,\ell=n-1)$ resulting in a strong K$\alpha$ transition $2p\rightarrow 1s$ whereas the other K transitions $np\rightarrow 1s$ are much weaker. The increasing importance of collisional processes with density leads to the decrease of the K$\alpha$ yield and simultaneously the increase of the other K$i$ yields leaving the absolute total yield practically unchanged. 
 
Experimentally, only relative K X-ray yields have been reported in the density range $\varphi \approx 10^{-7}-10^{-1}$\,\cite{And84,Bre98,Lau98}. As a whole, for these relative yields the agreement between theoretical results and experimental data is very good practically for all densities under consideration\,\cite{Jen02c,Pop11,LLM16}. Comparison of theoretical results\,\cite{Pop11} with experimental data \,\cite{And84,Bre98} suggests a modified statistical initial $\ell$-distribution at low densities (below $\varphi \approx 10^{-4}$). At higher densities, the initial $\ell$-distribution is reshuffled immediately because of Stark mixing. Combining the theoretical information\,\cite{LLM16} of the absolute yield $Y_{\rm K\alpha}+Y_{\rm K\beta}=83.7\%$ and the experimental result\,\cite{Lau98} for the $Y_{\rm K\alpha}/Y_{\rm K\beta}$ ratio of $1.46\,\pm\,0.25$ leads to an absolute yield $Y_{\rm K\beta}=(34\,\pm\,3)$\% of the K$\beta$ transition at a target density $\varphi = 10^{-2}$, which is in very good agreement to the theoretical prediction\,\cite{LLM16} $Y_{\rm K\beta}=33.3\%$ as given in fig.\,\ref{figure:fig3_K_yield_absolute}. 

Besides a highly desirable measurement of absolute K X-ray yields with high statistics, in particular in the range $\varphi = 10^{-3}- 1$, the high resolution studies of the Doppler broadening yield insight in the relative importance of the Coulomb de-excitation.

\subsection{Natural line width of the {$\bm{\mu$}}H K{$\bm{\beta$}} transition}

The natural line width of the atomic transition $3p\rightarrow 1s$ is practically given by the lifetime of the $3p$ state. Here, the $3p$-level radiative width of 20\,$\mu$eV contributes most. An estimate for the induced width from Stark mixing $3p\leftrightarrow 3s$ and external Auger effect for the $3p$ state, based on transition rates for $\pi^-p$ given in ref.\,\cite{Jen02a}, yields a negligibly small contribution of $\leq 1\,\mu$eV at target densities $\varphi \sim 10^{-2}$. 

Collisional broadening amounts at maximum to the magnitude of the natural line width being about 10$\mu$eV, and a maximal contribution of 3\,meV is due to thermal motion. In the absence of any Doppler broadening, the measured line shape would be defined by the resolution function of the spectrometer (see sec.\,\ref{subsec:resolution}).

\subsection{Kinetic energy distributions}

\begin{figure}[h]
\begin{center}
\resizebox{0.5\textwidth}{!}{\includegraphics{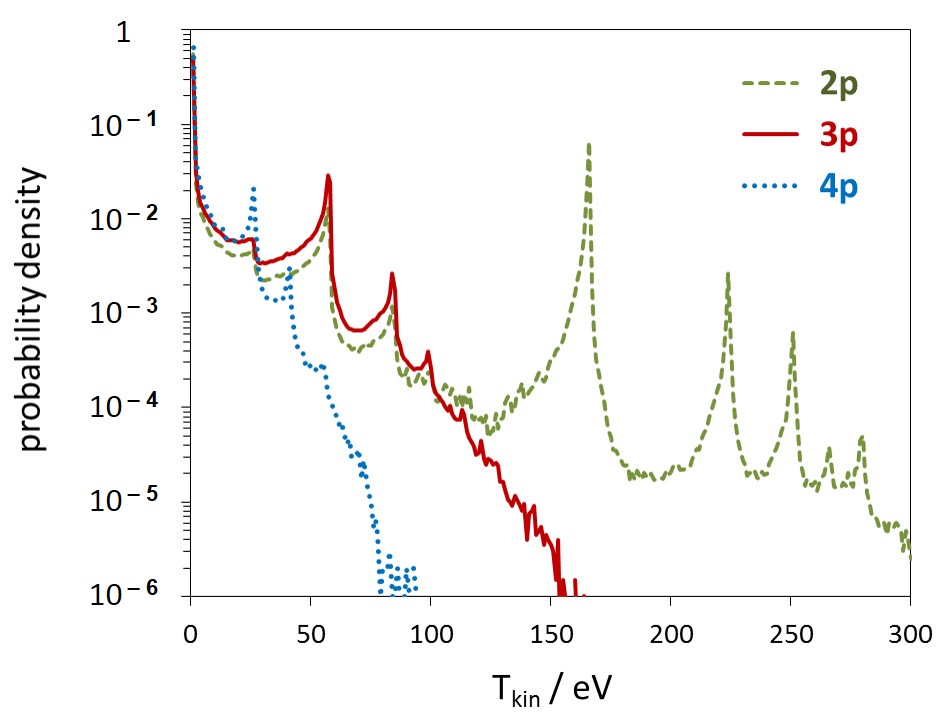}}
\caption{Kinetic energy distributions of $\mu^{-}p$ atoms at the instant of the radiative $np \rightarrow 1s$ transitions for $n=2,3$, and 4 in gaseous hydrogen at a density equivalent to a pressure of $10$\,bar ($\varphi\sim 1.5\cdot 10^{-2}$).}
\label{figure:fig4_Tkin_K}
\end{center}
\end{figure}

The kinetic energy distribution of the exotic atom changes during the cascade.  Studying its shape depending on individual atomic states allows a more stringent test of the cascade models than the X-ray yields alone. The $\mu^{-}p$ kinetic energy distributions at the instant of the $np\rightarrow 1s$ radiative transitions ($n=2-4$), as calculated with the cascade model described above, for a target pressure of $10$\,bar ($\varphi\sim 1.5\cdot 10^{-2}$) and room temperature  are shown in fig.\,\ref{figure:fig4_Tkin_K}. 

The calculated kinetic energy distributions have a distinctive high-energy structure arising from the various $\mathrm{\Delta}n\geq 1$ Coulomb transitions preceding the radiative de-excitation. The complicated shape of these structures evolves from the kinetic energy distribution before the Coulomb transition, the anisotropy of the angular distribution in the Coulomb de-excitation process and, finally, the deceleration due to elastic scattering and Stark transitions after the Coulomb de-excitation.

\begin{figure}[h]
\begin{center}
\resizebox{0.49\textwidth}{!}{\includegraphics{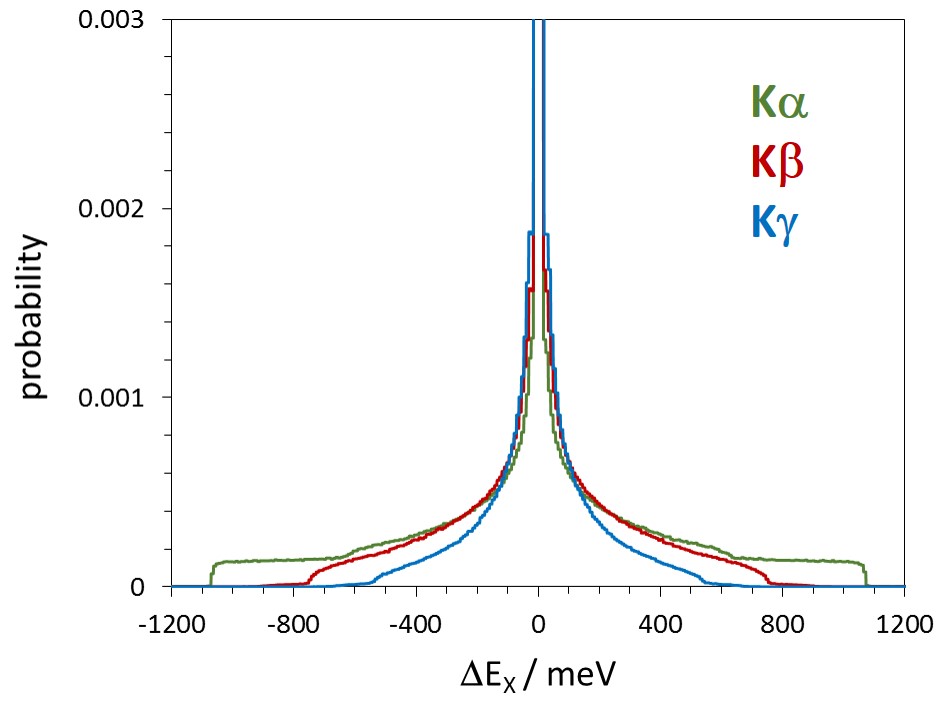}}
\caption{Effect on the X-ray line shape for various K transitions in muonic hydrogen induced by the kinetic energy distributions as shown in fig.\,\ref{figure:fig4_Tkin_K}. All distributions are generated with $4\cdot 10^{6}$ events and normalised to one. The narrow central peaks representing the low-energetic part of the kinetic energy distribution are cut to illustrate the increasing tails with decreasing initial principle quantum number $n$. Maximum value of the central peak is 0.0148, 0.0153, and 0.0183 for the K$\alpha$, K$\beta$, and K$\gamma$ transition, respectively.} 
\label{figure:fig5_Doppler_Kbeta}
\end{center}
\end{figure}

To demonstrate the effect of Coulomb de-excitation, various X-ray line shapes are displayed in fig.\,\ref{figure:fig5_Doppler_Kbeta} for an ideal apparatus, {\it i.\,e.} without the spectrometer resolution taken into account. The Doppler-induced broadening manifests itself in particular in the tails of the line. A reliable detection of such a broadening requires a well controlled and low background in the experiment.

\subsection{Molecular formation}\label{subsec:molform}

It is known from muon-catalysed fusion experiments that metastable hybrid molecules like $[(pp\mu)p]ee$ are formed during $\mu$H + H$_2$ collisions\,\cite{Taq89,Jon99}. The observation of fast $(\mu^{-}p)_{1s}$ atoms was taken as evidence for such resonant molecular formation in $2s$ excited states\,\cite{Poh01,Poh09}. For $\mu p(2s)$ quenching, however, the origin of the fast $(\mu^{-}p)_{1s}$ atoms either via resonant formation or via direct Coulomb de-excitation in $\mu p+$H collisions is still under debate\,\cite{Pop11,Die13}.

Such complexes are assumed to stabilise non-radiatively by Auger emission. Possible X-ray transitions from weakly bound excited molecular states before stabilisation would falsify the value for the hadronic shift determined from the measured X-ray energy due to satellite lines. Such transitions would be shifted only slightly to lower energies, whereas Auger stabilized molecules emit X-rays of at least 30\,eV energy less than the undisturbed transition\,\cite{Kil04} and are easily resolved in this experiment (see  discussion in sec.\,\ref{subsec:spectrum}). Radiative decay of such molecular states has been discussed and its probability is predicted to increase with atomic mass\,\cite{Kil04,Lin03}.


\section{Experimental approach}\label{sec:exp}

\subsection{Set-up of X-ray source}\label{sec:source}

\begin{figure}[b]
\begin{center}
\resizebox{0.485\textwidth}{!}{\includegraphics{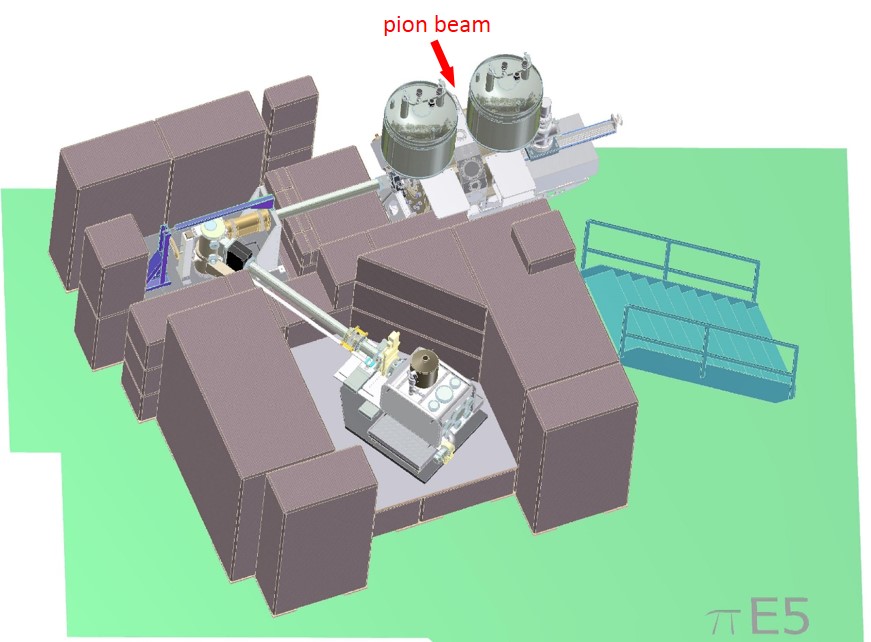}}
\caption{Set-up for the muonic hydrogen experiment in the $\pi$E5 area at the Paul Scherrer Institut (PSI). A massive concrete shielding is essential to suppress beam-induced background. The roof of the shielding is omitted\,to\,show\,the vacuum\,system connecting  cyclotron trap (upper right), crystal chamber (upper left) and the cryostat of the X-ray detector (bottom left). Details of the mechanical set-up are shown in figs.\,\ref{figure:cyc-cry} and \ref{figure:cry-det}.}
\label{figure:setup_muH}
\end{center}
\end{figure}

The experiment was performed at the $\pi$E5 channel of the proton accelerator at PSI (fig.\,\ref{figure:setup_muH}), which provides negatively charged pions with intensities of up to a few $10^{8}$/s. The pion beam adjusted to a momentum of 112\,MeV/c was injected into the cyclotron trap\,II \cite{Sim8893}, which consists of a superconducting split-coil magnet separated by a wide warm gap of 30\,cm. Inside the gap, a vacuum chamber is inserted, where the pions were decelerated by using a set of polyethylene degraders. The particles are guided by the weakly focusing magnetic field, being perpendicular to the beam direction, in spirals towards the magnet's centre, where a hydrogen filled cylindrical gas cell was installed along the symmetry axis of the magnet (fig.\,\ref{figure:cyc-cry}).  

Because of the short life time, slowing down of  pions must occur within a few turns. Muonic hydrogen is formed from the fraction of muons originating from the decay of almost stopped pions close to the cell. Such muons can be  trapped in the bottle field of the trap, oscillate there, and finally form muonic atoms.

The hydrogen cell of 220\,mm length together with the wide gap allows slow muons to spiral through the gas without being captured at the cryostat walls of the magnet. For a target density equivalent to about 10\,bar and with a degrader set optimised on the number of muon stops, from about 0.2\% of the incoming pions the decay muons form muonic hydrogen atoms. The degrader optimization was performed by measuring X-rays from muonic helium with a fast read-out charge-coupled device (CCD) having a sensitive area of 1\,cm$^2$ and a pixel size of 150\,$\mu$m\,\cite{Gor05}. The CCD was placed about 70\,cm away from the target cell in the bore of the magnet, where later on also the crystal spectrometer was attached (see figs.\,\ref{figure:setup_muH} and \ref{figure:cyc-cry}).

One end cap of the cell having 58\,mm in diameter is closed by a mylar$^{\textregistered}$ foil of 5\,$\mu$m thickness to keep absorption losses small for the X-rays. The window foil is supported by 1\,mm thick horizontal aluminium bars having a 6\,mm spacing, which can be operated safely at 1\,bar gas pressure. The opening aperture is given by a $40\times 40$\,mm$^2$ aluminium mask. The hydrogen gas was kept at a pressure of $(1.01\,\pm\,0.01)$\,bar and cooled to a temperature of $(25.5\,\pm\,0.2)$ K by means of a cold finger in order to enhance the stop density. The target temperature was  stabilized by means of a cooling feedback loop. In this way, the gas density was kept at $\varphi = (1.45\pm 0.30)$\%, which corresponds to a pressure of $(11.3\pm 0.3)$\,bar at $s.t.p.$ (=\,1\,bar at 273\,K).

\subsection{Crystal spectrometer}\label{sec:spectrometer}

Reflection-type crystal spectrometers in  Johann geometry allow the simultaneous measurement  of a complete energy interval according to the width of the X-ray source in the direction of dispersion. A corresponding detector extension is needed as well. Such a set-up requires bending of the crystal in the diffraction plane. In this experiment, spherically bent crystals were used which, due to a partial vertical focussing, additionally enhance the spectrometer efficiency. Details on the imaging properties of such a spectrometer can be found in ref.\,\cite{Got16}.

At the density $\varphi\approx 1.5\%$, the K$\beta$ line yield is maximal though being a factor of about 2 lower than the one of the K$\alpha$ line (fig.\,\ref{figure:fig3_K_yield_absolute}). At the K$\alpha$ energy of 1.9\,keV, however, only quartz is available for ultimate energy-resolution spectroscopy. Considering the reflectivity of the quartz (100) at 1.90\,keV and the silicon (111) reflection at 2.25\,keV, together with absorption losses in detector and target windows, yields a figure of merit in favour of the Si (111)/K$\beta$ combination by a factor of about 4. Furthermore, in silicon secondary reflections do not exist for the wavelength of the $\mu$H K$\beta$ line.

The silicon plate of 0.3\,mm thickness was attached by molecular forces to a glass lens polished to optical quality, which essentially determines the curvature. The bending radius was measured to be $R=(2982.2\pm 0.3)$\,mm. The crystal's reflecting area of 10\,cm in diameter was restricted by a circular aperture to 90\,mm in diameter in order to avoid edge effects. No horizontal limitation is needed because at a Bragg angle of 61$^{\circ}$ and for the large bending radius (tab.\,\ref{table:setmuH}) Johann broadening ---the leading aberration effect--- is negligibly small\,\cite{Egg65,Zsc82}. 

\begin{figure}[b]
\begin{center}
\hspace*{-1mm}
\resizebox{0.49\textwidth}{!}{\includegraphics[angle=0]{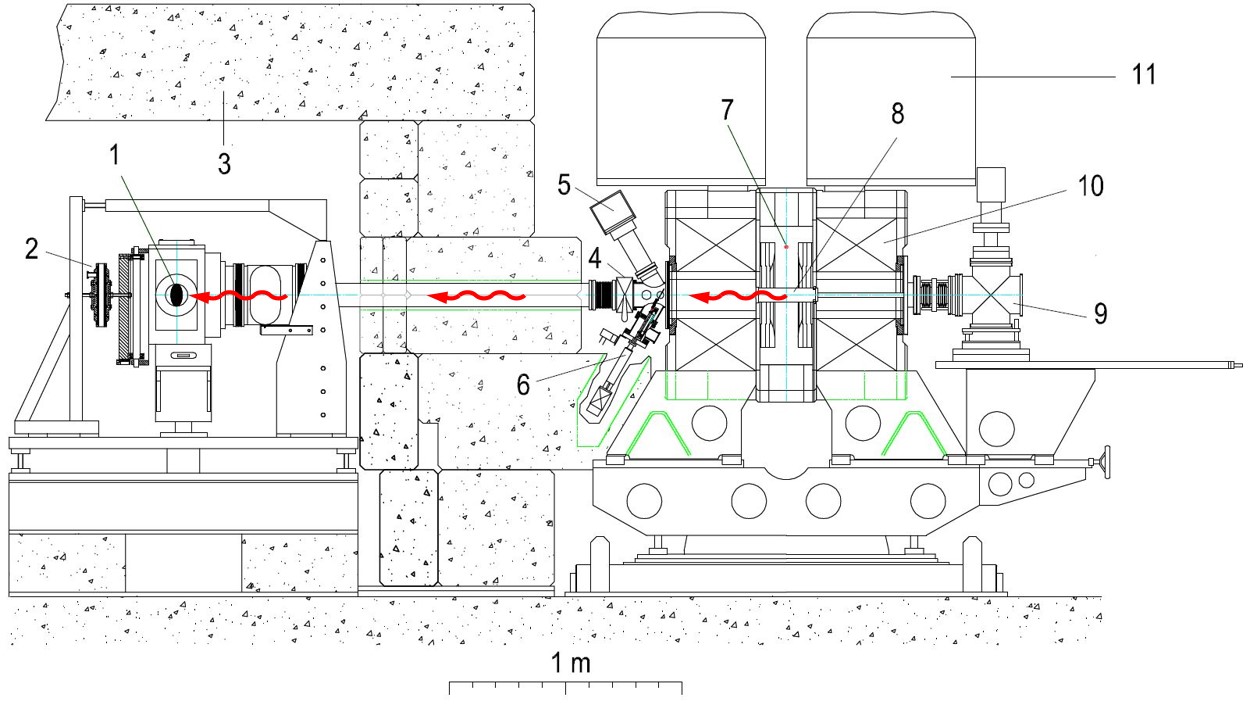}}
\caption  {Sectional drawing of the connection of the cyclotron trap to the crystal chamber of the spectrometer.  
The symmetry axis of the magnetic field coincides with the line centre--of--the--trap to centre--of--Bragg crystal.  Wavy arrows indicate the X-ray path from the hydrogen cell or the selenium fluorescence target towards the Bragg cystal. 1: Bragg crystal, 2: traction relaxation, 3: concrete shielding, 4: gate valve, 5: X-ray tube, 6: retractable selenium fluorescence target, 7: position of pion beam without B field, 8: target cell, 9: target cell support and cryogenic generator, 10: magnet coils of cyclotron trap, 11: liquid helium dewar.}
\label{figure:cyc-cry}
\end{center}
\end{figure}

Being pixel detectors, charge-coupled devices are ideal detectors for X-rays in the few keV range, because they combine an intrinsic position resolution with the good energy resolution of semiconductor detectors. In this expe\-riment, a 3\,$\times$\,2 array of CCDs was used covering a total sensitive area of 72\,mm in height and 48\,mm in width\,\cite{Nel02}. The individual devices of $24\times$24\,mm$^2$ consist of 600$\times$600 pixels of 40$\times$40\,$\mu$m$^2$. With an open electrode structure on the CCD's surface, a quantum efficiency is achieved of 85\% at 2.25\,keV\,\cite{Hol96}. The cryostat containing the CCD array is separated from the spectrometer vacuum by a 5\,$\mu$m thick mylar$^{\textregistered}$ foil.

The relative orientation of the individual devices as well as the pixel size at operation temperature of $-100^{\circ}$C was obtained by means of an optical measurement with a nanometric grid\,\cite{Ind06}. The detector surface was oriented perpendicular to the direction crystal -- detector. 

\begin{figure}[b]
\begin{center}
\hspace{-1mm}
\resizebox{0.49\textwidth}{!}{\includegraphics[angle=0]{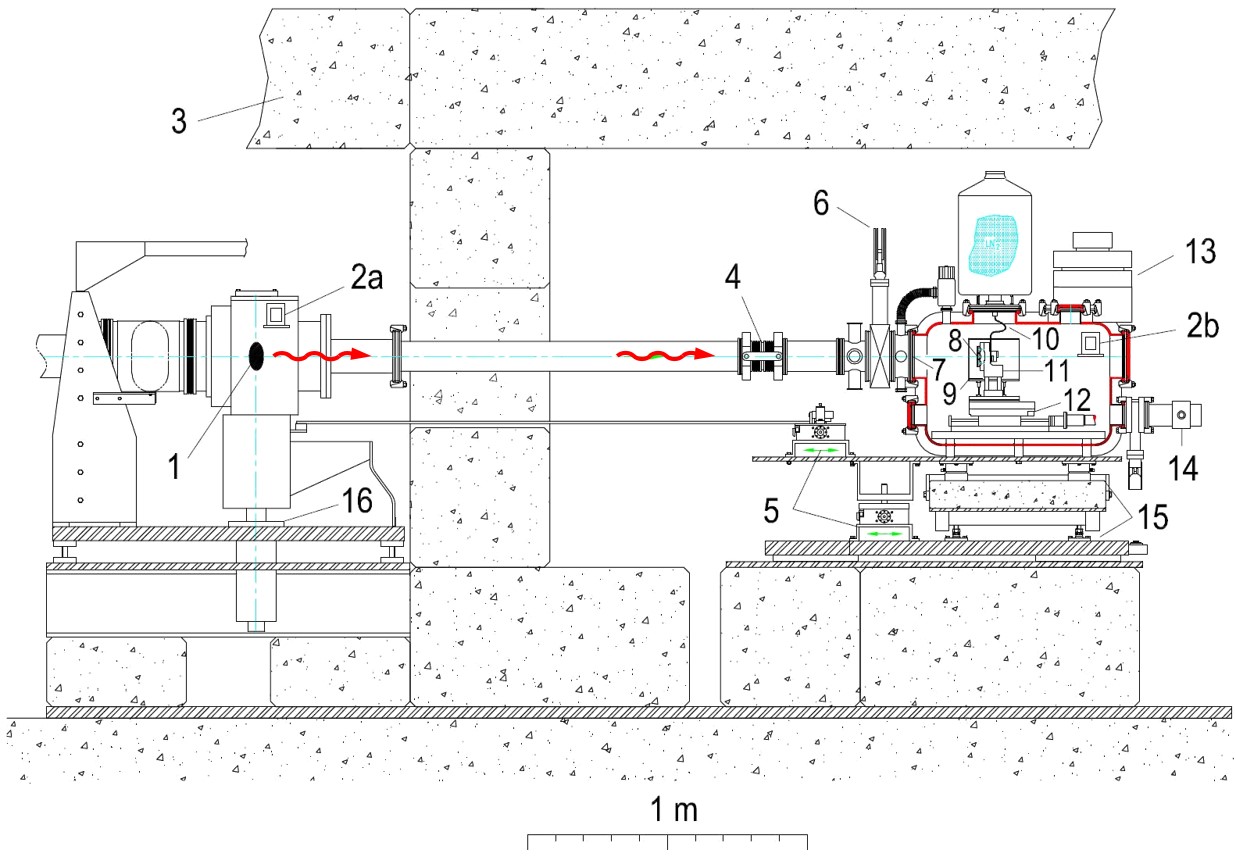}}
\caption {Sectional drawing of the connection of the crystal chamber to the X-ray detector. Wavy arrows indicate the X-ray path from the Bragg crystal towards the detector. 1: Bragg crystal, 2a and 2b: inclination sensors, 3: concrete shielding, 4: compensator, 5: linear tables for crystal and detector adjustment, 6: gate valve, 7: 5~$\mu $m Mylar window, 8: CCD detector array, 9: cold trap, 10: copper braid, 11: cold finger, 12: translation table for focal length adjustment, 13: detector readout electronics, 14:  turbomolecular pump, 15: air cushion support, 16: crystal chamber support.}
\label{figure:cry-det}
\end{center}
\end{figure}

The nominal focal position is given by the focusing condition $R\cdot\sin\mathrm{\Theta_{B}}=2621.7$\,mm for the horizontal plane. It takes into account a miscut angle of $(0.228\pm 0.006)^{\circ}$ and its orientation relative to the symmetry plane of $(-87.4\pm 1.2)^{\circ}$ as found from a dedicated measurement\,\cite{Cov08}. The distance crystal -- detector, chosen to be at the assumed K$\beta$ focal length (fig.\,\ref{figure:cry-det}), was found to be $(2622.4\pm 0.3)$\,mm from a survey measurement. 

The distance from the centre of the crystal to the centre of the cyclotron trap was chosen to 2620\,mm (fig.\,\ref{figure:cyc-cry}). Thus, the window of the target cell is about 5\% inside the Rowland circle given by the focal condition $R\cdot\sin\mathrm{\Theta_{B}}$. 

Monte-Carlo studies show that about 90\% of the reflection is covered by the height of the CCD array. The solid angle of the crystal with respect to the cell window is approximately $9\cdot 10^{-4}$\,sr. The  fraction of the X-ray source covered by the crystal's angular acceptance is approximately $1.5\%$ for the given target aperture. Taking into account window absorption and assuming a peak reflectivity of 40\% for the Bragg crystal\,\cite{San98}, the overall efficiency of the spectrometer results in about $4\cdot 10^{-7}$.

\setlength{\tabcolsep}{0.4mm}
\begin{table}
 \caption[Setup data PSI]
   {X-ray energies and set-up parameters for the crystal spectrometer. Energies of the hyperfine transitions are obtained from the calculated spin-averaged $\mu^{-}p$ transition energy\,\,\cite{Indpc} and the ground-state hyperfine splitting of (182.7$\,\pm\,$0.1)\,meV\,\cite{Mar04}. The splitting of the $3p$ level is ignored (see sec.\,\ref{sec:analysis}). The Se K$\alpha$ energies were taken from \,\cite{Des03}. The Bragg angles $\mathrm{\Theta_{B}}$ are obtained by using 2d$_{111}$=(0.62712016\,$\pm$\, 0.00000001)\,nm\,\cite{Bas94} for twice the lattice distance. $n$ is the order of reflection, $\mathrm{\Delta\Theta}_{\mathrm{IRS}}$ the angular shift owing to the index of refraction, and $y_{CD}$ the calculated focal length for the corresponding 
energy.}
  \label{table:setmuH}
\begin{tabular}{lrclccccc}
 \hline \\[-3mm]
                 & \multicolumn{3}{c}{energy}  &$n$& $\mathrm{\Theta_{B}}$  & $\mathrm{\Delta\Theta}_{\mathrm{IRS}}$ &$y_{\mathrm{CD}}$ \\
                 & \multicolumn{3}{c}{/\,eV}    &   &                        &         & /\,mm   \\[1.0mm]
 \hline\\ [-3mm]
 $\mu$H$(3p-1s)^{\,1}S_0$ & \multicolumn{3}{c}{2249.598} & 1\, & $61^{\circ}30'54.9"$   &  44.4"  & 2621.6  \\[1.0mm]
 $\mu$H$(3p-1s)$          & 2249.461& $\pm$& 0.001       & 1\, & $61^{\circ}31'18.1"$   &  44.4"  & 2621.7  \\[1.0mm]
 $\mu$H$(3p-1s)^{\,3}S_1$ & \multicolumn{3}{c}{2249.416} & 1\, & $61^{\circ}31'25.8"$   &  44.4"  & 2621.8  \\[1.0mm]
 Se K$\alpha_1$           & 11222.52 & $\pm$& 0.12       & 5\, & $61^{\circ}44'38.1"$ &  1.9"   & 2627.2  \\[1.0mm]
 Se K$\alpha_2$           & 11181.53 & $\pm$& 0.31       & 5\, & $62^{\circ}8'14.1"$  &  2.0"   & 2636.8  \\[1.0mm]
  \hline
  \end{tabular}
\end{table}

The stability of the set-up was monitored twofold. First, the peak position was used of the Se K$\alpha_1$ fluorescence radiation excited by means of an X-ray tube. The Se and $\mu$H measurements were performed alternating at least once per day. For this, selenium powder was attached over an area of $28\times 25$\,mm$^2$ on a pure aluminium support placed retractable remote-controlled in a distance of about 2.1\,m from the crystal (fig.\,\ref{figure:cyc-cry}). In one Se measurement, typically 25000 K$\alpha_1$ events were recorded. 

The advantage of placing the X-ray source somewhat off the target-sided focal position is an averaging over possible non uniformities of the fluorescence material. As long as the angular acceptance of the crystal has full overlap with the target for the energy interval needed no suppression in the tails of the reflections occur. Both the Se target and the window of the H$_2$ target cell were extended enough in the direction of dispersion to avoid such cuts. 

Second, two inclination sensors (of 1\,$\mu$rad precision) measured the relative orientation of crystal and detector in 15\,min periods. One sensor was attached to the crystal chamber and the second one to the detector cryostat, where the relative positions of crystal and detector to the respective chamber are mechanically fixed. Though in principle tightly connected by the vacuum system, crystal and detector are mounted on individual supports (Fig. 8). The inclination sensors allow to detect possible relative movements of the vertical axes of crystal and detector housings, which would result in a line broadening because of the tilt of the reflection. The relative movements have been found to by negligibly small.

In this way, both short- and long-term movements of the vertical crystal axis relative to the X-ray detector were recognized. The maximal deviations from long-term movements were found to be about $\pm$\,0.1\,mrad corresponding to about $\pm\,0.5$ pixel as well from the Se control measurements and the inclination sensors. Larger short-term amplitudes due to liquid nitrogen refilling of the detector ($1-2$ minutes twice per day) are easily identified by the inclination sensors. The corresponding periods were excluded from the analysis. Almost 11000 events in the $\mu$H K$\beta$  line were recorded in a three weeks measurement.

\subsection{Data processing}\label{subsec:spectrum}

Raw data recorded by the detector system  consist of the pixel's digitized charge contents and a position index. The penetration depth of 2.25\,keV X-rays in silicon is 2.1\,$\mu$m \cite{Vei73} and photo-electron ranges are below 100\,nm. Consequently, all X-rays are converted in the surface structure or in the depletion layer of the CCDs, where charge diffusion is small. Therefore, the charge originating from the conversion of these X-rays is deposited in one or two pixels with one common boundary. Beam induced background, mainly high energetic photons from neutrons produced in pion  absorption and captured in surrounding nuclei, lead to larger structures and can be recognised by a pattern analysis. Hence, the granularity of CCDs allows for efficient background rejection by such a cluster analysis. Defect pixels of CCDs are masked by software. 

\begin{figure}[t]
\begin{center}
\resizebox{0.4\textwidth}{!}{\includegraphics{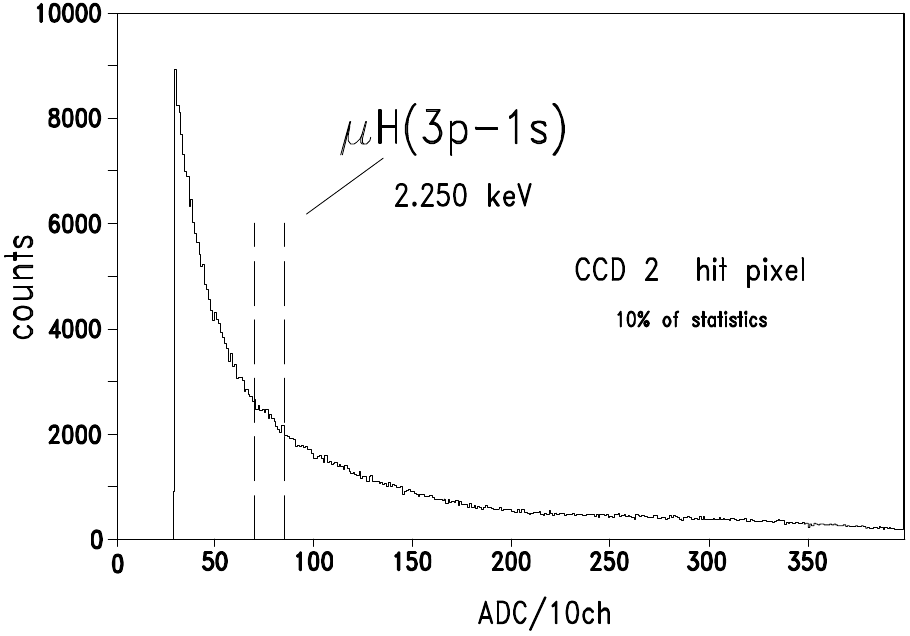}}
\resizebox{0.4\textwidth}{!}{\includegraphics{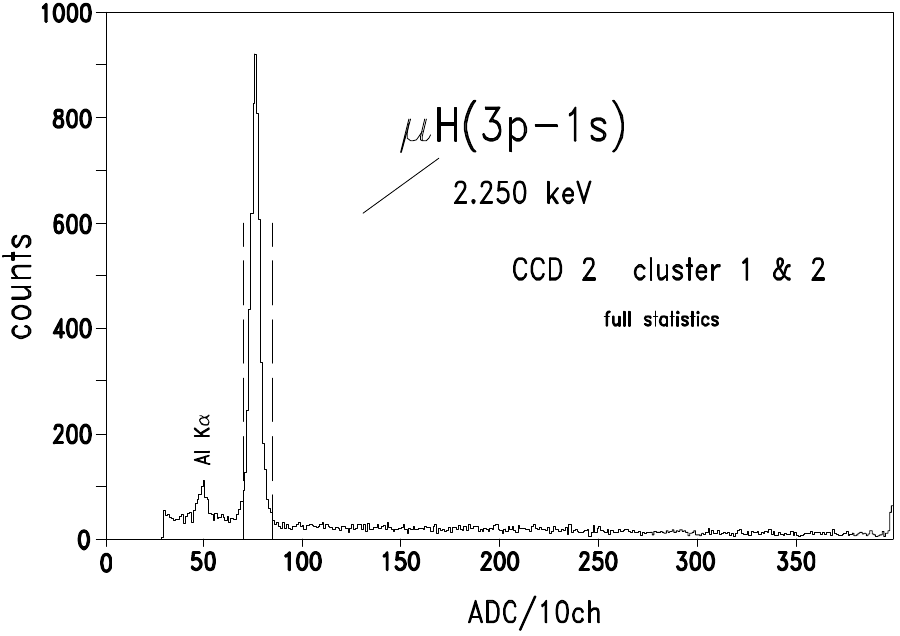}}
\caption{Spectrum of the collected (digitized) charge in the centre CCD of the right column containing the $\mu$H$(3p-1s)$ reflection (see fig.\,\ref{figure:scat_muHline}) before (top) and after (bottom) cluster analysis (10 ADC channels are compressed to one). The peak around the (compressed) channel 80 is due to the K$\beta$ transition energy of 2.25\,keV. Vertical dashed lines indicate the applied "energy cut" of $\pm\,3\sigma$ allowing for further background reduction.}
\label{figure:ADC_CCD}
\end{center}
\end{figure}

Inspecting cluster sizes reveals ---as expected--- that mainly single ($\approx$\,70\%) and two pixel events ($\approx$\,28\%) contribute. The spectra of the collected charge cleaned in this way show a pronounced peak originating from the 2.25\,keV muonic hydrogen K$\beta$ X-rays (fig.\,\ref{figure:ADC_CCD}). No difference in the X-ray line shape is identifiable for the single and the single plus two pixel event spectra within statistics.

\begin{figure}[t]
\begin{center}\hspace*{-2mm}
\resizebox{0.5\textwidth}{!}{\includegraphics[angle=0]{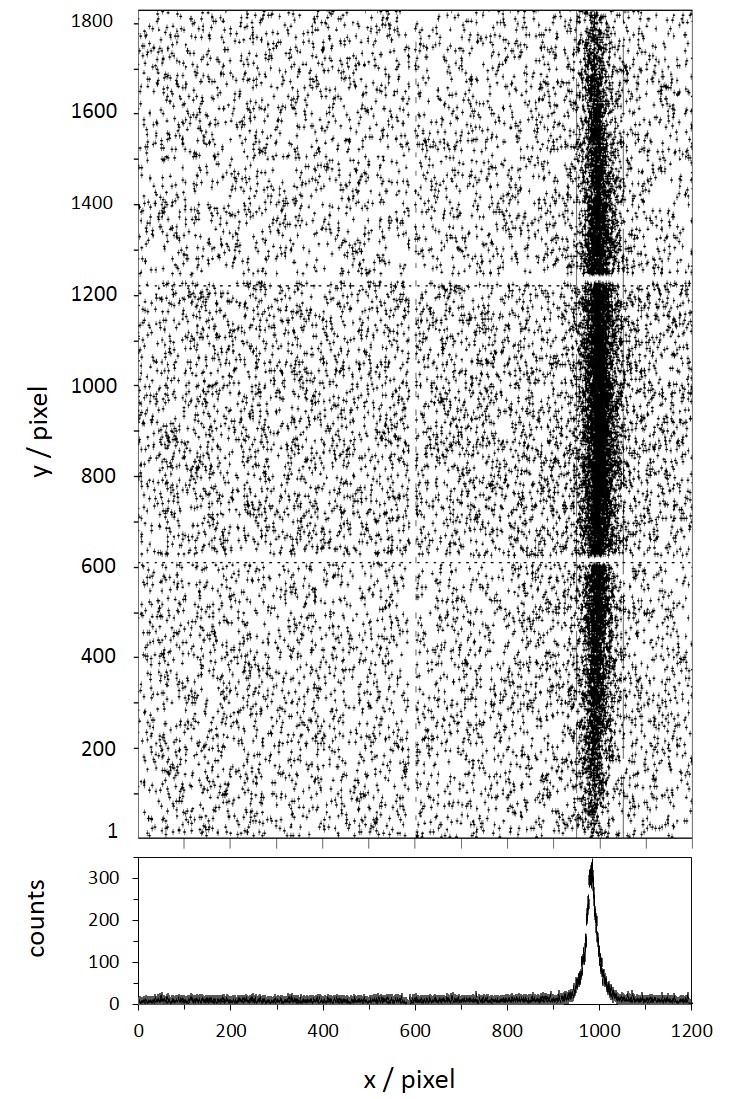}}
\caption{Scatter plot and projection to the axis of dispersion (after curvature correction) for the $\mu$H$(3p-1s)$ reflection. It covers the 3 CCDs of the right column of detector. The curvature parameters are obtained from a parabola fit. One channel (equal to one CCD pixel) corresponds to $(18.616\pm0.003)$\,meV in the direction of dispersion ($x$) for first order reflection.}
\label{figure:scat_muHline}
\end{center}
\end{figure}

The detector resolution in terms of the collected charge is determined by means of a Gaussian fit to the K$\beta$ peak and found to be 170\,eV (FWHM). By applying a cut on the charge with an interval corresponding to $\pm\,3\sigma$, an additional significant background reduction is achieved in the two-dimensional position spectra (fig.\,\ref{figure:scat_muHline}). 

The total measuring time was subdivided into 3 one week periods. Beginning of each period, for the typical accelerator current of 1.8\,µA, the K$\beta$ count rate was about 40 per hour dropping to 25 per hour within the week. This decrease is attributed to an ice layer on the cryogenic target increasing with time.  Therefore, the target was warmed up to room temperature before each measuring period.

In principle, for energies as low as 2.2\,keV the change of the transmission through mylar windows and ice layer over the line width must be considered. However, the crystal's also energy dependent reflectivity partly compensates the effect of the absorption. The difference, taking into account windows and maximum ice layer, over an energy interval of 1\,eV results in about $2\cdot10^{-4}$, which is negligible in view of the accumulated statistics. 

The hit pattern of the X-rays on the CCD surface shows a curvature originating from the imaging properties of the reflection geometry. The curvature is corrected by means of a parabola fit before projecting onto the axis of dispersion, which is equivalent to an energy axis. The uncertainty of the curvature correction influences the line width by less than 0.1 pixels or 2\,meV which is negligible in view of the total line width of almost 700\,meV.

The position of the X-ray pattern was chosen such as to allow the measurement of an  interval of almost 20\,eV towards lower energies. A search for molecular states, which had been discussed in sec.\,\ref{subsec:molform}, lead to an upper limit for satellite X-ray transitions from Auger stabilised states  of 1\% (3$\sigma$) in this energy range (fig.\,\ref{figure:scat_muHline}).

\subsection{Spectrometer resolution}\label{subsec:resolution}

The spectrometer response was measured at the energies 3104, 2765, and 2430\,eV using the narrow M1 X-ray lines from helium-like argon, chlorine, and sulphur produced in a dedicated electron-cyclotron ion resonance trap (ECRIT)\,\cite{Bir00}. With the same geometrical set-up as used for the experiment, the hydrogen target was replaced by an extended X-ray source where the helium-like ions were  provided by an ECR discharge in the centre of the cyclotron trap. Originally designed to determine precisely the crystal's bending radii and resolutions for various apertures, the measurements were also used to identify the miscuts and their orientations\cite{Ana05,Tra07,Covth}. Furthermore, the results were extensively used to validate the X-ray tracking code which includes the aberrations to all orders for the assumed geometry. As a main result, the crystal response function was obtained for different energies with an accuracy of 1\% or better in terms of FWHM.

The resolution function at a given energy can be built up from the rocking curve as calculated from the dynamical theory of diffraction for a perfect flat crystal (intrinsic resolution) convoluted with the geometrical imaging by means of a Monte-Carlo ray-tracing code constituting the ideal response. The intrinsic resolution (rocking curve) is calculated here for the wavelength given with the code XOP\,\cite{San98}. For the muonic hydrogen K$\beta$ energy of 2.25\,keV, the rocking-curve width is found to be 245\,meV.

\begin{figure}[b]
\begin{center}
\hspace*{-4mm}
\resizebox{0.45\textwidth}{!}{\includegraphics[angle=0]{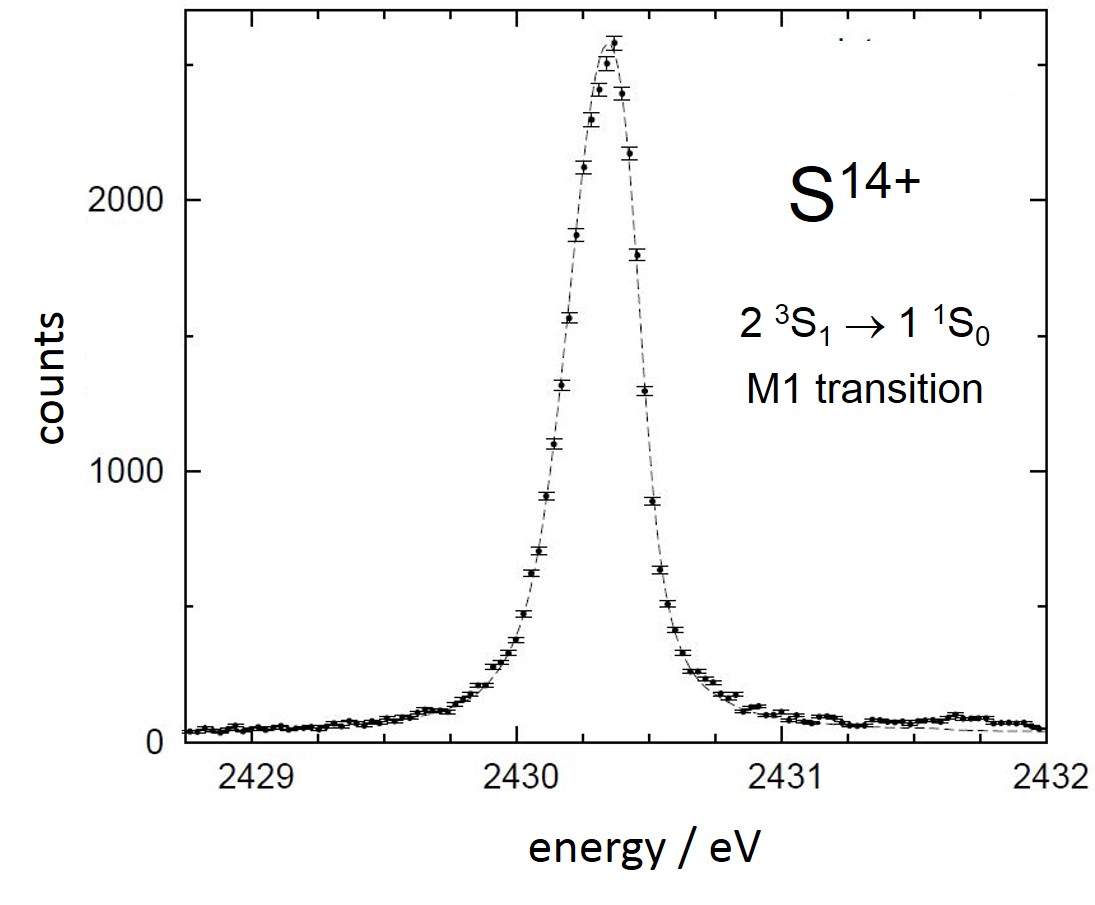}}
\caption{Spectrum of the M1 transition in helium-like sulphur used for the determination of the spectrometer resolution.}
\label{figure:S14+}
\end{center}
\end{figure}

An additional Gaussian contribution served modelling the 
imperfections of the crystal material and mounting. The value for the additional Gaussian width at the energy of the K$\beta$ transition energy was found from a fit to the results for argon, chlorine, and sulphur and extrapolation to 2.25\,keV to be ($92\,\pm\,4$)\,meV. Figure\,\ref{figure:S14+} shows the M1 transition of helium-like sulphur.

This method yields a width of the resolution function of (272$\,\pm$\,3)\,meV (FWHM), a value close to the theoretical limit of 254\,meV comprising the value of 245 meV for a plane crystal calculated by means of XOP and the broadening caused by the geometry of the experiment. The remaining difference is due to the additional Gaussian contribution. Details on the experimental set-up and analysis may be found elsewhere\,\cite{Cov08}.

\begin{figure}[h]
\begin{center}
\hspace*{-4mm}
\resizebox{0.5\textwidth}{!}{\includegraphics{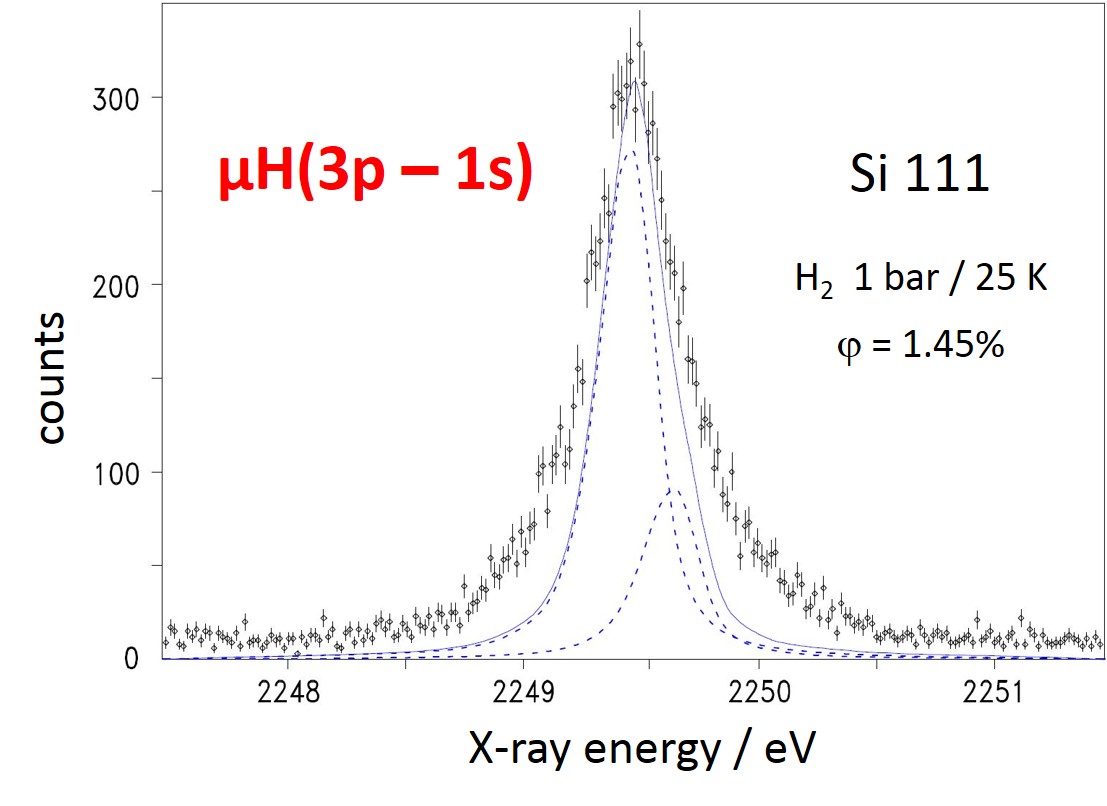}}
\caption{Measured line shape of the $\mu$H K$\beta$ transition $(3p-1s)$ compared to the spectrometer response (thin solid line). The superposition of the contributions of the individual hyperfine components (dashed lines) represents the line width if any Doppler broadening is absent (normalised to the peak height after background subtraction). The target parameters (1\,bar/25\,K) correspond to 1.45\% liquid hydrogen density. One energy bin corresponds to 2 CCD pixels or 37.2\,meV.}
\label{figure:fig11_muH3-1_b2_fit}
\end{center}
\end{figure}

In fig.\,\ref{figure:fig11_muH3-1_b2_fit}, the Monte-Carlo-generated response for the set-up is shown, where the total response consists of the superposition of the two hyperfine components. A comparison with the measured K$\beta$ line shape reveals a significant additional broadening which is attributed to the Doppler-induced contributions from Coulomb de-excitation. 

\section{Data analysis}
\label{sec:analysis}

\subsection{Approach to the line shape}
\label{subsec:approach}

The measured spectrum shown in fig.\,\ref{figure:fig11_muH3-1_b2_fit} has been fitted with the following parameter set:

\begin{itemize}
\item[1.] energy splitting of the hyperfine doublet $\mathrm{\Delta} E_{\mathrm{hfs}}$,
\item[2.] relative intensity of the hyperfine components $R_{\mathrm{T/S}}$,
\item[3.] total intensity,
\item[4.] absolute line position,
\item[5.] background level (assumed to be constant), and
\item[6.] kinetic energy distribution causing the Doppler broadening.
\end{itemize}

For the physics case considered here, only the quantities corresponding to the last-mentioned item are of interest. In continuation of the analysis as described in ref.\,\cite{Cov09}, we describe  different approaches to determine the kinetic energy distribution of the $\mu^{-}p$ system from the observed line 
broadening.

Nonetheless, parameters 1 and 2, the splitting and population of the $1s$ hyperfine levels, provide an important  consistency check for all approaches applied in the analysis. The theoretical $1s$ hyperfine splitting of 182.7\,meV\,\cite{Mar04} is confirmed by the result of a measurement of the $2s$ splitting\,\cite{Ant13} when using the corresponding scaling law including corrections\,\cite{Mar05,Pes16}. For the relative population of the $^3S_1$ and $^1S_0$ states, we do not expect any deviation from the statistical distribution of 3:1.

The level splittings of the $3p$ state are about 2\,meV\,\cite{Pac96}, which is about two orders of magnitude smaller than the width of the spectrometer resolution function of 272\,meV (see sec.\,\ref{subsec:resolution}). Thus, the effects of the $3p$ hyperfine splitting (4 lines instead of 2 forming the transition $3p \to 1s$) are too small to have any significant influence on the shape, and they are neglected in the further analysis.

The kinetic energy distribution has been treated in three different ways:
\begin{itemize} 
\item[1.] Inspired from the discrete energy release owing to  presumably strong low-lying Coulomb de-excitation transitions, the kinetic energy distribution was modelled by a few narrow intervals of typically a few eV width. Fitting of position and relative intensities of the different components is referred below as phenomenological
approach. Up to three such components with their intensities summing up to 100\% were used in this analysis.
\item[2.] The data were compared to a kinetic energy distribution, like the one displayed in fig.\,\ref{figure:fig4_Tkin_K}, which is the result of a calculation using the most recent cascade theory\,\cite{Pop11}. 
\item[3.] A deconvolution procedure was developed as a method to extract information about a kinetic energy distribution of the $\mu^{-}p$ atom in the $3p$ state directly from the data.
\end{itemize}

\subsection{Fitting and deconvolution techniques}
\label{subsec:techniques}

Decisive evaluation criteria in data analysis are, among others, that any procedure should include as well an error estimate and the assessment of the quality of the evaluation (goodness of fit). A survey of different kinds of  data evaluation for high energy physics problems is given in a recent review by G. Cowan \cite{Ber12}. Two main approaches are presented there, the so-called frequentist and the Bayes method. Both methods have their merits which have been  discussed  in numerous publications, {\it e.\,g.} in \cite{Cou95,Ber12,Efr86,Jam03,Boh10,Tra17} and references therein. Far from contributing to this discussion, we tried to use the advantages of each of the methods when they seemed appropriate.

As a starting point, we re-evaluate the phenomenological model used in ref.\,\cite{Cov09}. While revealing the basic features of the Doppler broadening, the analysis described in ref.\,\cite{Cov09} was based on the frequentist method  and had serious shortcomings related to error estimates
and unresolved correlations.

In order to circumvent some of the above-mentioned problems, the phenomenological approach was reinvestigated. Such a comparison is favourably done by using the Bayesian framework and, consequently, used extensively for this purpose. In order to demonstrate the gain of knowledge from the Bayesian over the frequentist method, which is in particular a realistic assessment of the uncertainties, the results of the frequentist approach described in \cite{Cov09,Cov09a,Covth} are summarised in sec.\,\ref{subsec:fit}. The results of the Bayesian approach are given in detail in sec.\,\ref{subsec:Bayes_results}.

The Bayesian method was also chosen for a direct comparison of the measured spectrum with theoretical predictions. As the kinetic energy distribution is parameter free  in such a study, the extraction of the triplet-to-singlet ratio and the hyperfine splitting constitutes an important consistency check.

For the sake of completeness, the most important features for the presented analyses are given in app.\,A for both the frequentist and the Bayesian method. The deconvolution technique, described in app.\,B1, also uses the Bayesian method. The result of the deconvolution is given in sec.\,\ref{subsec:deconv}.


\section{Analysis of the line shape}\label{sec:results}

\subsection{Results from the {\bm{$\chi^2$}} (frequentist) analysis}\label{subsec:fit}

The phenomenological approach assumes a low-energy component ranging up to a few eV and a few high-energy components as suggested by the distribution shown in fig.\,\ref{figure:fig4_Tkin_K}. The most important high-energetic components are expected to be at about 58 and 27\,eV owing to the Coulomb de-excitation transitions $(4-3)$ and $(5-4)$. 

The spectra constructed in this way are compared to data by means of a $\chi^2$ analysis using the MINUIT package\,\cite{Jam75}. Hyperfine splitting, relative intensity of the two lines, background, and relative weight of the Doppler contributions are free parameters of the fit. 

\begin{figure}[t]
\resizebox{0.45\textwidth}{!}{\includegraphics{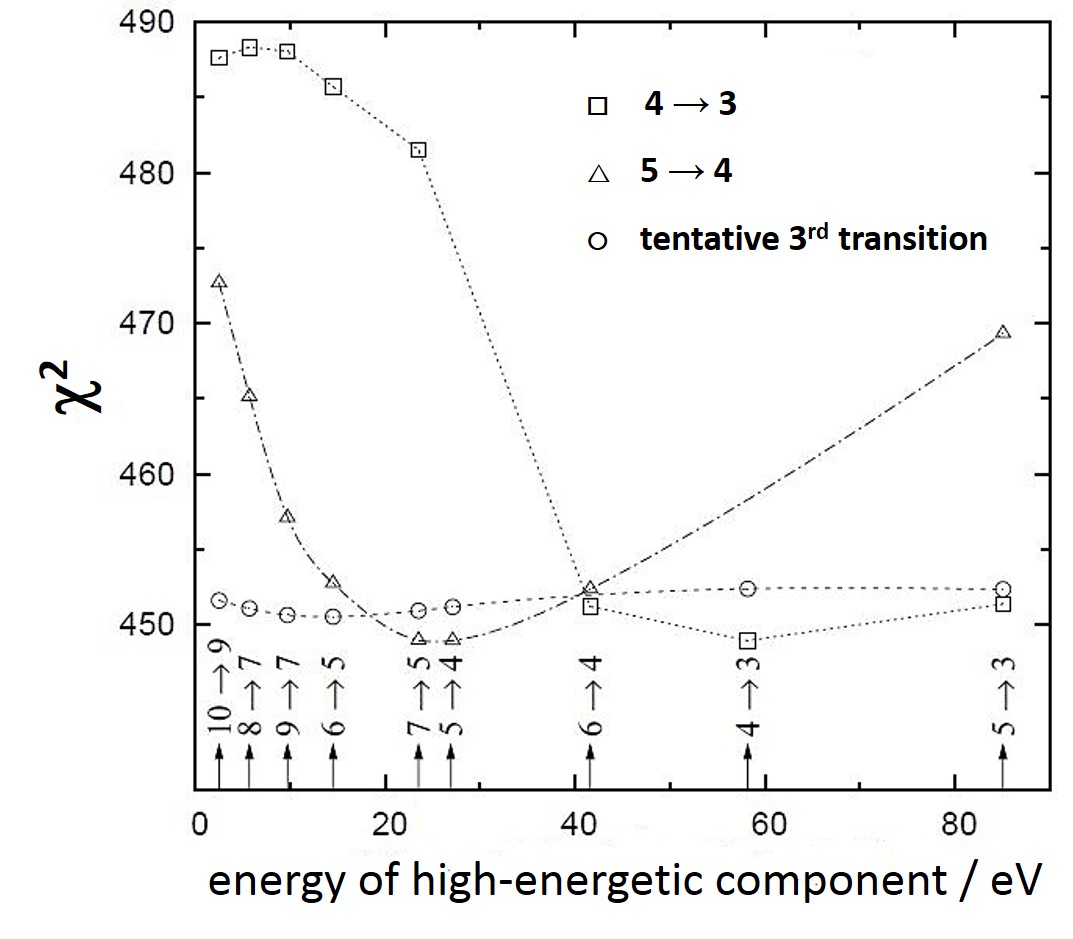}}
\caption{Result of the $\chi^2$ (frequentist) analysis modeling the kinetic energy distribution with 3 narrow energy intervals. Symbols indicate the tentative assignment to a Coulomb de-excitation transition. More details may be found in\,\cite{Covth}.}
\label{fig12_3box_fit} 
\end{figure}

The optimum $\chi^2$ could be achieved by using two high-energetic contributions ($\chi^2_{\mathrm{min}}=448.9$ with a number of degrees of freedom of $n.d.f=474$) and found at $T_{\mathrm{kin}}\approx 24$ and $T_{\mathrm{kin}}\approx 58$\,eV (fig.\,\ref{fig12_3box_fit}). These energies correspond indeed to the $(5-4)$ and $(4-3)$ Coulomb de-excitation transitions. However, the component at 58\,eV is identified only when the component at the position 24\,eV is assumed to exist, where the fit is widely insensitive to the widths of these components. In a second iteration, where the components at 24\,eV and 58\,eV are assumed to exist, the best range of the low-energy component was then found to be $T_{\mathrm{kin}}=0-2$\,eV. A fit of almost the same quality is achieved with one component only at an energy of about 40\,eV. 

No improvement is obtained by including a continuum of decelerated $\mu^{-}p$ atoms by broad uniform distributions from $2-56$\,eV and $58-88$\,eV between the discrete  energies of possibly important Coulomb transition, or when adding more narrow components. In particular, no hint for a $(5-3)$ transitions at 85\,eV could be found. 

For the best fit, the relative contributions of the kinetic energy components are found to be ($61\,\pm\,2$)\%, ($25\,\pm\,3$)\%, and ($14\,\pm\,4$)\% for the energy intervals [$0-2$]\,eV, [$26.4-27.4$]\,eV, and [$57.7-58.7$]\,eV, respectively. The optimum $\chi^2$ yields ($211\,\pm\,19$)\,meV and ($3.59\,\pm\,0.59$) for the hyperfine splitting and for the relative population of the ground state hyperfine states $^3S_1$ and $^1S_0$, respectively. Fixing the hyperfine splitting to the theoretical value of 182.7\,meV\,\cite{Mar04} leads to ($2.90\,\pm\,0.21$) for the triplet-to-singlet ratio, which is very close to the statistical value, with practically unchanged $\chi^2$. 

However, the (rather small) errors are due to a particular fit only and do not allow to draw any conclusions for systematic uncertainties. As discussed above, the fit does not reveal any further detail of the structure of the kinetic energy distribution.
 
Nevertheless, the following decisive conclusions remain: (i) the result of about 60\% for the intensity of the low-energy component is very robust with respect to the fit model, and (ii) high-energetic components, without being sensitive to details, are essential because an alternative broadening of the low-energy component alone yields always a large $\chi^2$.

\subsection{Results from the Bayesian analysis}
\label{subsec:Bayes_results}

The following describes results for the  phenomenological approach as obtained by means of the Bayes method. Within this approach, the evaluation of the evidence $\mathcal{Z}$ is central to the present work. Ratios of the evidences (Bayes factors) are used to assess which  description supports best the data (see app.\,A2). 

In order to handle such a multi-dimensional problem, the recently developed method of nested sampling was applied\,\cite{Ski04}. This method uses an efficient way to calculate the many-dimensional integral of the evidence by converting it into a one-dimensional integral. Moreover, it is able to calculate  probability distributions for individual and correlated parameters (posterior distributions) with minimal additional effort.

\subsubsection{Ground-state splitting and population}
\label{subsec:HFS}

\begin{figure}[h]
\begin{center}
\hspace*{-2mm}
\resizebox{0.51\textwidth}{!}{\includegraphics{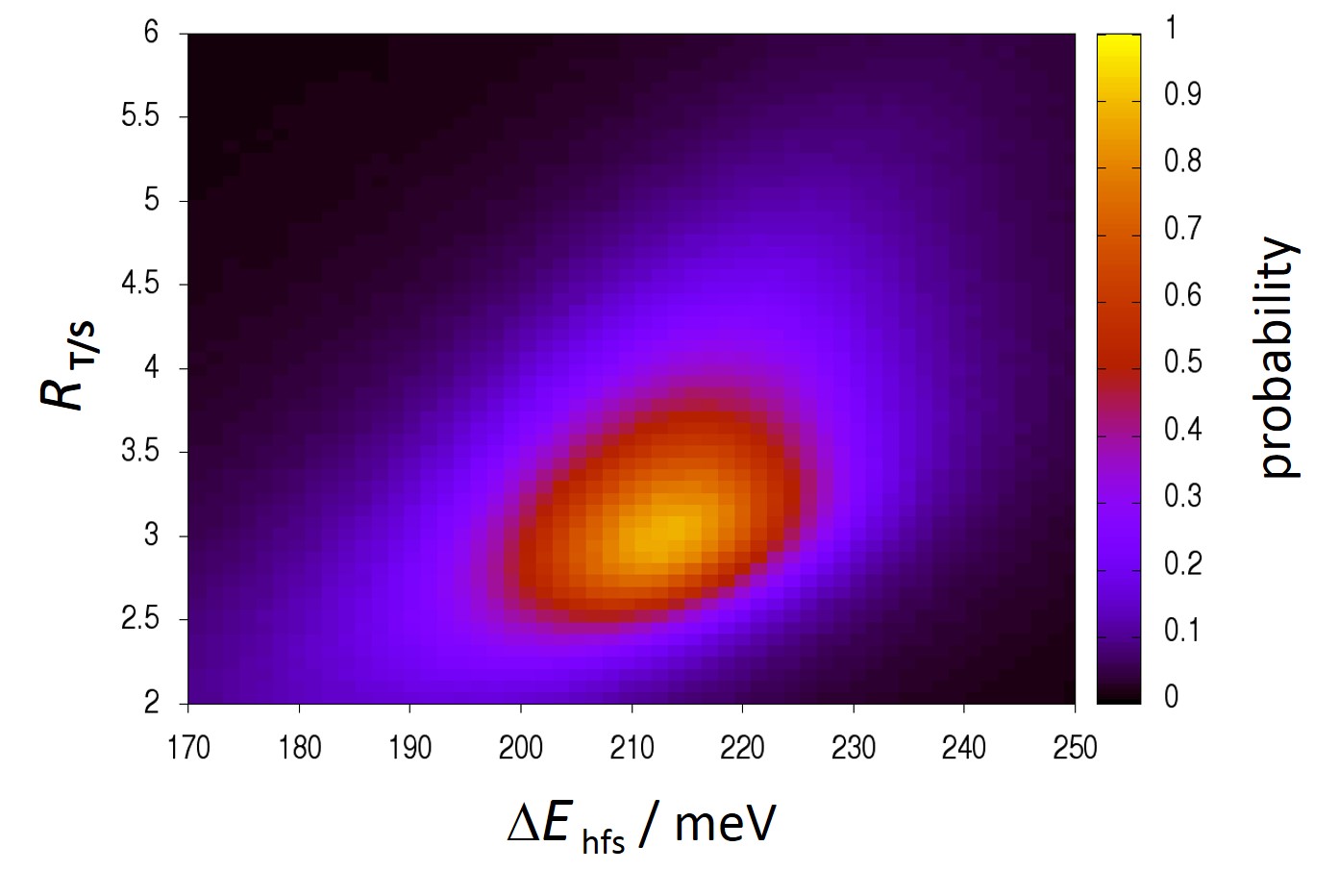}}
\caption{Contour plot for the correlation of hyperfine splitting $\mathrm{\Delta} E_{\mathrm{hfs}}$ and relative population $R_{\mathrm{T/S}}$ in a three component phenomenological model (see sec. 4.1) as obtained with the Bayesian method. The maximum probability is set to 1.}
\label{fig13_3box-hfs_G75} 
\end{center}
\end{figure}

The contour plot shown in fig.\,\ref{fig13_3box-hfs_G75} can  be compared to the equivalent one as obtained in the $\chi^2$ analysis (fig.\,3 in \cite{Cov09}). The results are consistent with the $\chi^2$ analysis\,\cite{Cov09} and are summarized in tab.\,\ref{table:results}. 

In the frequentist and the Bayesian analysis, $\mathrm{\Delta}E_{\mathrm{hfs}}$ is larger than the theoretical value by about 2$\sigma$ where $R_{\mathrm{T/S}}$ is in good agreement with the statistical value of 3 (the 1$^{\mathrm{st}}$ row in tab.\,\ref{table:results}). In view of the fact, that the two hyperfine components cannot be resolved, this result is consistent with the expectations.

\setlength{\tabcolsep}{2.1mm}
\begin{table}[b]
 \begin{center}
 \caption {Results from the Bayesian analysis assuming one low-energy (LE) and two high-energy components (phenomenological approach) and when using a fixed  kinetic energy distribution from an ab initio $ESCM$  calculation (cascade theory). Credibility limits are 1$\sigma$.}
  \label{table:results}
\begin{tabular}{clcrlcrl}
 \hline \\[-2mm]
model & \multicolumn{2}{c}{$\mathrm{\Delta}E_{\mathrm{hfs}}$ / meV } & \multicolumn{2}{c}{$R_{\mathrm{T/S}}$} & \multicolumn{2}{c}{intensity LE} \\
 \hline\\[-2mm]
phenom.\,app.& 212~ & \hspace*{-4mm}${+\,23}\atop{-\,22}$  & ~~3.2 & \hspace*{-4mm}${+\,1.6}\atop{-\,0.7}$& ~~~0.65 & \hspace*{-4mm}${+\,0.03}\atop{-\,0.04}$\\
 [3.0mm]
phenom.\,app.& 182.7 & \hspace*{-4mm}\,QED                  & ~~2.5 & \hspace*{-4mm}${+\,1.1}\atop{-\,0.5}$& ~~~0.64 & \hspace*{-4mm}${+\,0.03}\atop{-\,0.04}$\\[3mm]

cascade theory& 202~ & \hspace*{-4mm}${+\,11}\atop{-\,10}$                    & ~~2.3 & \hspace*{-4mm}${+\,0.5}\atop{-\,0.4}$& \multicolumn{2}{c}{---}\\[3mm]
cascade theory& 182.7 & \hspace*{-4mm}\,QED                  & ~~1.7 & \hspace*{-4mm}${+\,0.4}\atop{-\,0.3}$&\multicolumn{2}{c}{---}\\[1mm]
 \hline
  \end{tabular}
 \end{center}
\end{table}

\begin{figure}[h]
\begin{center}
\resizebox{0.5\textwidth}{!}{\includegraphics{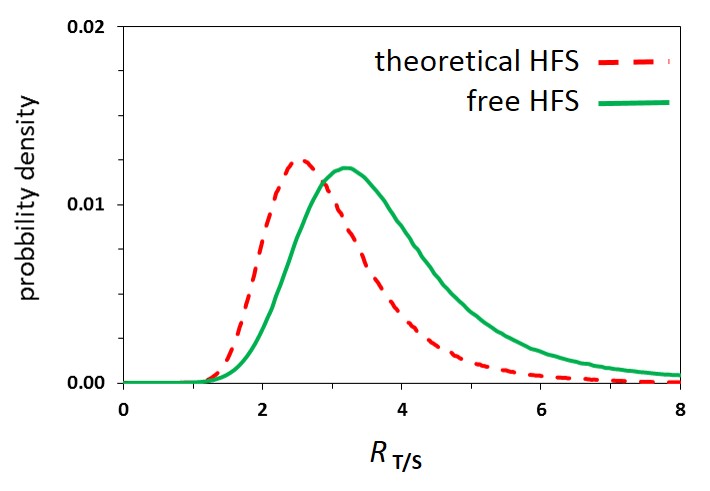}}
\caption{Comparison of the model-averaged intensity ratio $R_{\mathrm{T/S}}$ of the three-component analysis with hyperfine splitting (HFS) fixed to the theoretical value of 182.7\,meV and being a free parameter.}
\label{figure:fig14_T-to-S_ratio} 
\end{center}
\end{figure}

\begin{figure}[h]
\begin{center}
\resizebox{0.5\textwidth}{!}{\includegraphics{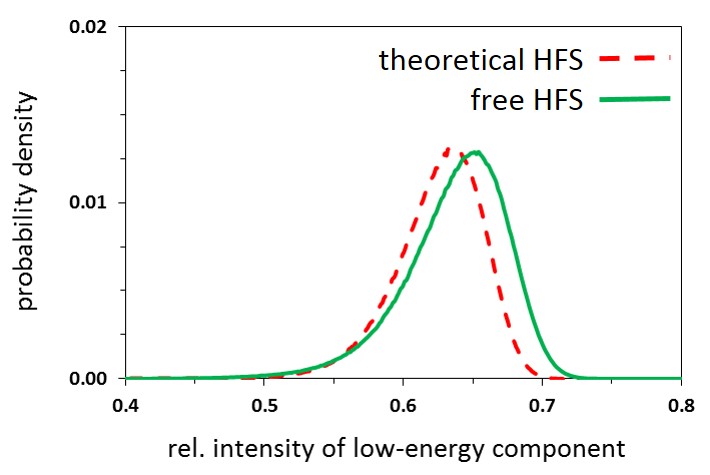}}
\caption{Comparison of the model-averaged relative intensity of the low-energy component in the 3-component analysis with fixed and free hyperfine splitting (HFS).}
\label{figure:fig15_Int_LE_component} 
\end{center}
\end{figure}

As can be seen also from fig.\,\ref{fig13_3box-hfs_G75}, the ratio $R_{\mathrm{T/S}}$ strongly correlates with the splitting $\mathrm{\Delta} E_{\mathrm{hfs}}$. Fixing $\mathrm{\Delta} E_{\mathrm{hfs}}$ to the calculated value leads to a decrease of $R_{\mathrm{T/S}}$ as expected but with credibility limits covering still well the statistical value of 3 (the 2$^{\mathrm{nd}}$ row in tab.\,\ref{table:results}). Figure\,\ref{figure:fig14_T-to-S_ratio} displays the corresponding probability distributions for  $R_{\mathrm{T/S}}$ and Fig.\,\ref{figure:fig15_Int_LE_component} for the relative intensity of the low energy component.

\subsubsection{Search for high-energetic components}
\label{subsubsec:HE_components}

As can be seen from table\,\ref{table:evidences_phenomenological}, a single component ---despite allowing any width--- is discarded immediately by the Bayes factor of $B\approx10^{\,26}$ when comparing to any approach including high-energetic components. Even more, it excludes any non-structured kinetic energy distribution at higher energies.

A 3-component approach is slightly favoured over the model with 2 components. Noteworthy to mention that according to Jeffreys' scale\,\cite{Jef61}, Bayes factors of 8 (20) are considered as weak (positive) evidence in favour of a model as opposed to another one corresponding to a credibility limit of 2$\sigma$ (3$\sigma$).

\begin{figure}[h]
\begin{center}
\resizebox{0.5\textwidth}{!}{\includegraphics{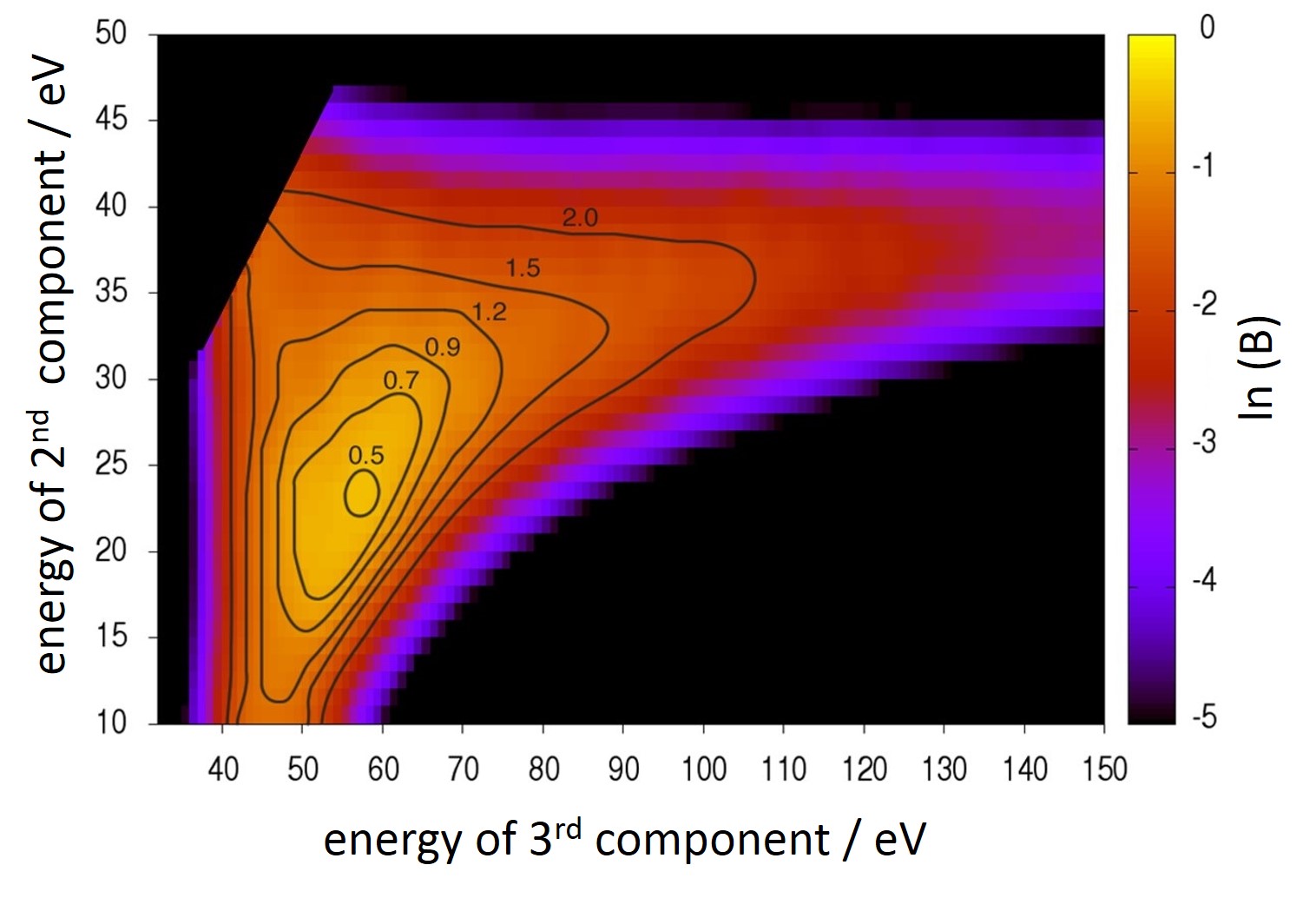}}
\caption{Search for two high-energy components using the Bayesian method. The most likely configuration  is found for 24 and 56\,eV as centres of 2\,eV wide intervals. Contour lines are log likelihood contours attributed according to the so called "Jeffreys' scale"\,\cite{Jef61,Kas95}. The result coincides with the $\chi^2$ analysis (fig.\,\ref{fig12_3box_fit}) but reveals the much larger uncertainties when performing an unbiased analysis. The cut on the upper left is due to the symmetric treatment of the two high-energy components yielding doubled statistics when overlayed.}
\label{fig16_3box_evidence} 
\end{center}
\end{figure}

The marginal distribution in the parameter space of a second and third component at higher energies is shown in fig.\,\ref{fig16_3box_evidence} with the hyperfine splitting being a free parameter. The low-energy part was fixed at the interval [$0-4$]\,eV and the widths of the two high-energy components were set to 2\,eV, because just as in the frequentist approach the analysis is insensitive to the detailed interval widths\,\cite{Cov09}.  

\setlength{\tabcolsep}{3mm}
\begin{table}[b]
 \begin{center}
 \caption {Comparison of phenomenological approaches to the kinetic energy distribution ordered by Bayes factors $B$.  Evidence is defined in app.\,A.2.1 eq.\,(\ref{eq:Z}). More details can be found in \,\cite{The13}.}
  \label{table:evidences_phenomenological}
\begin{tabular}{cccc}
 \hline \\[-3mm]
 model      & hyperfine     & maximum  & $B$ \\
            &  splitting    & evidence &     \\
 \hline\\[-2mm]
3-component  &  182.7\,meV  &  50231.4 &   1 \\[1mm]
3-component  &  free        &  50229.2 &   9 \\[1mm]
2-component  &  free        &  50228.4 &  20 \\[1mm]
1-component  &  free        &  50171.4 & $10^{26}$\\[1mm]
\hline
  \end{tabular}
 \end{center}
\end{table}  

In the following, the intervals of the second and third component were set to [$23-25$]\,eV and [$55-57$]\,eV. When fixing in addition the hyperfine splitting to the theoretical value of 182.7\,eV, the evidence again improves ($B=9$). The relative intensities obtained from the corresponding probability distributions ($1\sigma$) are found to be $\left(63{{+3}\atop{-4}}\right)$\% ([$0-4$]\,eV), $\left(24{{+~4}\atop{-10}}\right)$\% ([$23-25$]\,eV), and $\left(13{{+10}\atop{-~4}}\right)$\% ([$55-57$]\,eV). 

As a side remark, the now larger errors agree with the conventional wisdom that the 3$\sigma$ errors from a frequentist analysis realistically correspond to 1$\sigma$ errors\,\cite{Jef61,Kas95}. The same behaviour can be assumed for the population ratio $R_{\mathrm{T/S}}$ and splitting $\mathrm{\Delta} E_{\mathrm{hfs}}$ and the good agreement can be regarded as purely accidental for the 1$\sigma$ limits. In addition, the uncertainties do not reflect systematic effects as revealed by the asymmetry of the bayesian credibility limits (see sec.\,\ref{subsec:fit} and ref.\,\cite{Cov09}).

Intensities for the two high-energy components are\linebreak strongly correlated, while the relative intensity of the low-energy component is again very stable against the details of its own shape and that of the high energetic components.

The results of the (bias free) Bayesian analysis are  consistent with the previous $\chi^2$ analysis\,\cite{Cov09},  however, show that the uncertainties for the intensities of the two high-energy components are much larger. As can be seen from fig.\,\ref{fig16_3box_evidence}, any detailed structure beyond the low-energy component can hardly be deduced from this analysis. But, finding maximum evidence for kinetic energies at the position of expectedly strong Coulomb transitions matches the picture from the atomic cascade.

It is worthwhile to mention that from the Bayesian analysis  evidence is found for exactly two distinct high-energetic components without forcing the energy of either component in the right range. This constitutes a major advantage of this method over the $\chi^2$ fit (see sec.\,\ref{subsec:fit}). 

\subsection{Comparison with cascade theory}

When using the calculated kinetic energy distribution\, \cite{Pop11} and comparing the Bayes factors to the ones in tab.\,\ref{table:evidences_phenomenological}, it is important to notice that only one (hyperfine splitting free) or no free parameter affecting the line shape is available. Given that the Bayes factors differ only by two with and without allowing to vary the hyperfine splitting (tab.\,\ref{table:evidences_theory}), we conclude that the two models are indistinguishable.

The intensities of the theoretical spectrum accumulated up the boundaries of the energy intervals of the three-component model amount to 58\% ([$0-4$\,eV]),  75\% ([$0-27$\,eV]), and  94\% ([$0-57$\,eV]).  This is in agreement with both the frequentist and the Bayesian result (secs.\,\ref{subsec:fit} and \ref{subsec:Bayes_results}) in view of the errors together with the fact, that components of an intensity of $\lesssim$\,10\% cannot be identified, and that thereby the normalisation condition excludes contributions above 57\,eV.

The triplet-to-singlet ratio, however, is only poorly reproduced when fixing the kinetic energy distribution (tab.\,\ref{table:results}). The reason for such a behaviour supposedly lies in the reduction of free parameters\,\cite{Kas95}. 

If two models have a different number of free parameters, they still can be compared by adjusting the $log$ of evidence by an offset\,\cite{Kas95}. Applying this correction leads to an evidence for the theoretical kinetic energy distribution similar to the one for the best phenomenological approach.

\setlength{\tabcolsep}{2.9mm}
\begin{table}[h]
 \begin{center}
 \caption {Comparison of the theoretical kinetic energy distribution. Comparing the Bayes factors to the ones in tab.\,\ref{table:evidences_phenomenological}, it is important to notice that only one ($\mathrm{\Delta}E_{\mathrm{hfs}}$ free) or no free parameter affecting the line shape is available.}
  \label{table:evidences_theory}
\begin{tabular}{cccc}
 \hline \\[-3mm]
 model      & $\mathrm{\Delta}E_{\mathrm{hfs}}$  & maximum evidence & $B$ \\
 \hline\\[-2mm]
cacade theory&  free        &  50227.7 &  40\\[1mm]
cacade theory&  182.7\,meV  &  50227.0 &  81\\[1mm]
\hline
  \end{tabular}
 \end{center}
\end{table}

\section{Deconvolution}\label{subsec:deconv}

The goal of the deconvolution procedure, as described in app.\,B1, is to get an alternative access to the structure of the high-energetic part of the $\mu^{-}p$ kinetic energy distribution at the instant of the radiative transition $3p\rightarrow1s$ directly from the measured K$\beta$ X-ray spectrum. The accuracy of such an approach is essentially limited by statistics. In order to reduce the uncertainties, hyperfine splitting and relative population  were fixed to their theoretical values.

The result of this procedure is a Doppler spectrum where the influence of the measuring process is strongly reduced and ideally even eliminated, and which is converted into a cumulative energy distribution for comparison with theoretical approaches. The usage of the cumulative distribution instead of the probability density has the advantage that the whole analysis does not require operations similar to numerical differentiation prone to numerical instabilities.

\subsection{Method}

The deconvolution is based on the Bayesian method exploiting the fact that the instrumental resolution of the detection system had been measured with high precision (see sec.\,\ref{subsec:resolution}). It makes use of the distributive law valid for the folding process. In contrast to the frequentist method, it constitutes a non-parametric approach (see, {\it e.\,g.} \cite{Siv06} chap.\,6).

The feasibility of this method has been tested with two simulated spectra as described in more detail in app.\,B2. The first test used a simple kinetic energy spectrum with three distinct energy components (fig.\,\ref{fig:dec4} -- a) that was folded with the spectrometer response discussed in sec.\,\ref{subsec:resolution} resulting in a (simulated) X-ray spectrum (fig.\,\ref{fig:dec4} -- b). Such a distribution corresponds to the experimental information, {\it i.\,e.} in this case the measured K$\beta$ spectrum (fig.\,\ref{figure:fig11_muH3-1_b2_fit}). The deconvolution  reproduced the input Doppler spectrum almost exactly apart from small deviations pointing to the difficulty of reproducing sharp edges in the kinetic energy distribution (fig.\,\ref{fig:dec4} -- a and c). 

In the second test we tried to approach real conditions as close as possible by using for the simulation a state of the art kinetic energy distribution from the cascade theory\,\cite{Pop11}. The successful deconvolution of the corresponding spectrum corroborated the soundness of the method (fig.\,\ref{fig:dec2}).

\subsection{Results}

\begin{figure}[t]
\begin{center}
\hspace*{-2mm}
\resizebox{0.35\textwidth}{!}
{\includegraphics{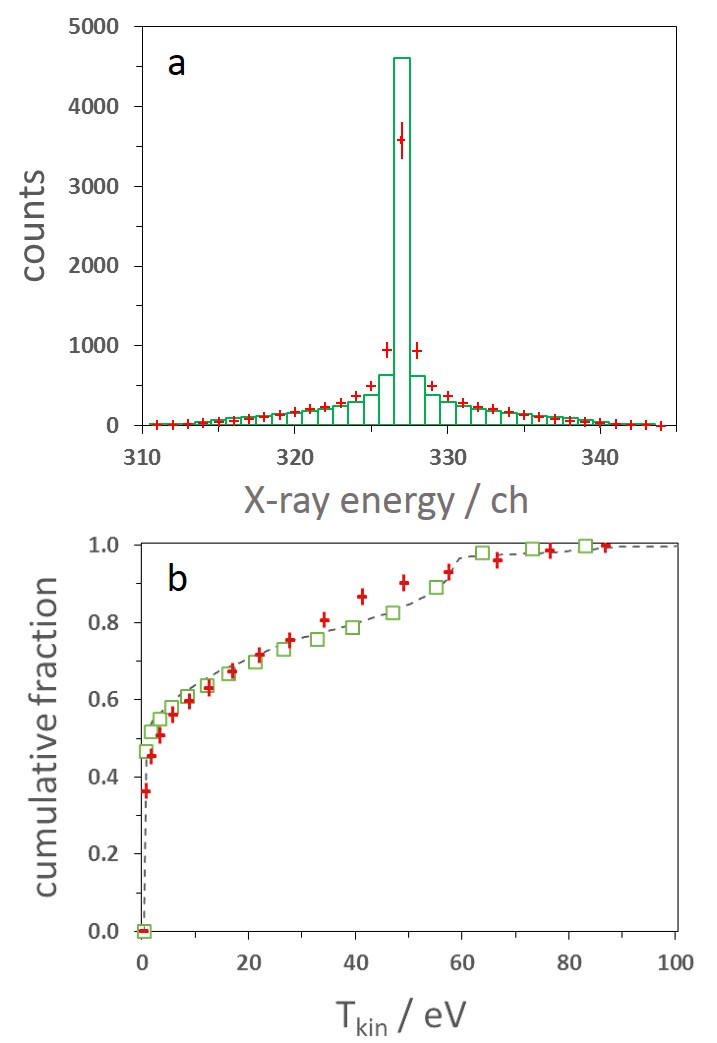}}
\caption{Deconvolution of the measured spectrum. a -- Reconstructed K$\beta$ X-ray Doppler broadening (points with error bars) in comparison with the Doppler spectrum as obtained from the kinetic energy distribution predicted by cascade theory\,\cite{Pop11} (histogram). b -- 
Cumulative energy distributions of the $\mu^{-}p(3p)$ state as reconstructed from the data (points with error bars), derived from the Doppler spectrum as predicted by cascade theory (squares), and the prediction of cascade theory (dashed line). Error bars represent the 1$\sigma$ level.}
\label{fig:dec3} 
\end{center}
\end{figure}

The Doppler spectrum resulting from the deconvolution of the measured X-ray spectrum is shown in fig.\,\ref{fig:dec3}  (points with error bars) in comparison with the prediction of the advanced cascade model\,\cite{Pop11} (histogram). The cumulative energy distributions derived from the corresponding\linebreak Doppler spectra are displayed in fig.\,\ref{fig:dec3} -- b. Points (with error bars) and squares are due to experiment and cascade theory, respectively. The cumulative energy  distribution as obtained directly from the calculated kinetic energy spectrum (fig.\,\ref{figure:fig4_Tkin_K}) is shown for comparison (dashed line).

At the 3$\sigma$ level, the reconstructed Doppler spectrum based on the data is in good agreement with the theoretical Doppler spectrum. However, there are a few distinct differences. The central (low-energy) peak of the reconstructed spectrum, has a smaller intensity and is broader than cascade  theory predicts. An attempt to explain such a broadening by a wrong response function used for the deconvolution was found to be inconsistent with the data. Therefore, we conclude that the resolution of such details is beyond the ability of the analysis the statistics given. 

These findings are also revealed in the cumulative distribution.  Again, there is some missing intensity at energies below 10\,eV in the reconstructed distribution. In the cascade calculation, 52\% of the intensity is found to be in the kinetic energy interval [$0-1$]\,eV, where the deconvolution yields (43\,$\pm$\,2)\%. As mentioned above, a 1\,eV wide low-energy component is not detectable in the phenomenological approach (see secs.\,\ref{subsec:fit} and \ref{subsubsec:HE_components}). For the interval [$0-4$]\,eV, the cumulated intensity in this analysis amounts to (54\,$\pm$\,2)\% to be compared with the result of the 3-component model of $\left(63{{+3}\atop{-4}}\right)$\%. The differences constitute about the limitation of accuracy achievable with the experimental data available for this analysis. All errors correspond to the 1$\sigma$ criterion.

More intensity is clearly seen in the reconstructed spectrum at medium energies around 40\,eV and again less intensity is found at high energies around 70\,eV. Due to the normalisation condition, any underestimation in the lower-energy part must be compensated at higher energies. The deviation, however, is at most only 3$\sigma$ for the complete spectrum.

In order to quantify these findings, a $\chi^{2}$ analysis has been performed by comparing the Doppler spectrum as obtained from the deconvolution of the data with the one based on the cascade theory yielding $ \chi^{2} / n.d.f.  = 9.3$ with $ n.d.f. =32 $. Varying the background within appropriate limits (see app.\,B1) did not yield a noticeable change for the resulting distribution.


\section{Summary}
\label{sec:summary}

The K$\beta$ X-ray transition in muonic hydrogen has been studied at a gas density of $\varphi = (1.45\pm 0.30)$\% of liquid hydrogen using a high-resolution crystal spectrometer. The main goal of this paper was to investigate the Doppler broadening originating from the kinetic energy distribution of the $\mu^{-}p$ atoms in the $3p$ state at the instant of the X-ray emission.  For the analysis of the X-ray line shape, both a frequentist ($\chi^2$) analysis and a Bayesian method were used. 

The shape of the K$\beta$ spectrum has been analyzed as a function of the ground-state hyperfine splitting, the corresponding triplet-to-singlet ratio, and the kinetic energy distribution in the $3p$ state. In a phenomenological approach, the kinetic energy distribution was modelled by three narrow energy intervals. The results for the energy splitting and the relative population of the ground-state hyperfine states are fully consistent with experimental data and QED calculations for the splitting and the expected statistical ratio $3:1$. 

Two high-energetic components corresponding to the expected Coulomb-de-excitation transitions ($5-4$) and ($4-3$) have been found in both the $\chi^2$ analysis and the Bayesian one. Both methods produce similar results for the intensities of the high-energy components; the Bayesian analysis, however, provides a better assessment of the models used for the data fit and the corresponding model parameters.    

A specially developed deconvolution method based on the Bayesian technique was used to determine the kinetic energy distribution directly from the shape of the K$\beta$ spectrum in a model independent way. In order to evaluate the efficiency of this method with a realistic kinetic energy distribution, we used the results of the recent advanced  cascade calculations \cite{Pop11} to simulate a measured  spectrum and then applied the deconvolution to reconstruct the Doppler broadened X-ray spectrum and from that a cumulative kinetic energy distribution. Taking into account uncertainties due to experiment and analysis, the reconstructed distribution was found to be in a good agreement with the theoretical input. The success of this self-consistency check allows us to expect that our method of the deconvolution of the experimental spectrum itself provides a reliable and model-independent way to determine the kinetic energy distribution of the $\mu^{-}p$ atoms at the instance of the K$\beta$ transition.   

The result of the deconvolution confirms the existence of the high-energy components found in the phenomenological approach. However, the deconvolution procedure has a clear advantage in comparison with the phenomenological approach since it is based only on the basic properties of the cumulative energy distribution (a monotonically increasing non-negative function) and does not involve any assumptions  about the shape of the energy distribution and the corresponding phenomenological parameters.    

The comparison of the reconstructed cumulative energy distribution of the $\mu^{-}p$ atoms in the $3p$ state with the results of the most-advanced cascade models \cite{Pop11} shows that there is a good qualitative agreement between the experiment and the theory. It constitutes the best theoretical description achieved so far. On the quantitative level, there are, however, still differences between the theory and the experiment both at low and high kinetic energies, which leave room for improvement.

The experimental spectrum analysed in this paper constitutes the present state of the art, and we are hardly able to extract more details reliably as given by the Bayesian analysis. To resolve remaining discrepancies in the Doppler broadening between the experimental data and the cascade calculations, one needs, in principle, new experiments with significantly better statistics, best together with an improved spectrometer resolution for different transitions preferably both in muonic hydrogen and deuterium.


\section*{Acknowledgements}
We are grateful for discussions on the hyperfine splitting in muonic hydrogen to A. Antognini and F. Kottmann and would like to thank N.\,Dolfus, B.\,Leoni, L.\,Stohwasser, and K.-P.\,Wieder for the technical assistance. The Bragg crystal was manufactured by Carl Zeiss AG, Oberkochen, Germany.
Partial funding is acknowledged to FCT, Lisbon (through LIBPhys – Unit 4559), the Germaine de Sta$\ddot{e}$l exchange program, and Grant Z1-WTZ: RUS11/A03. LKB is Unit\'{e} Mixte de Recherche du CNRS, de l'\'{E}cole Normale Sup\'{e}rieure et de UPMC No. 8552. This work is part of the PhD thesis of one of us (D.\,S.\,C., Univ. of Coimbra, 2008). 
\section*{Author contribution statement}
D. Gotta, V. E. Markushin, and L. M. Simons conceived the experiment strategy. D. Gotta and L. M. Simons conceived and planned the experiment setup and supervised the project. V. N. Pomerantsev and V. P. Popov developed the theoretical aspects of the atomic cascade and performed corresponding calculations which include contributions from V. E. Markushin and T. S. Jensen. D. S. Covita and L. M. Simons set up the cyclotron trap and D. Gotta the crystal spectrometer. H. Gorke designed, set up, and operated the CCD used for the degrader optimization and analysed the corresponding data with the help of M. Nekipelov. H. Fuhrmann, A. Gruber, A. Hirtl, T. Ishiwatari, Ph. Schmid, and J. Zmeskal designed, set up, and operated the cryogenic target. P. Indelicato, E.-O. Le Bigot, and M. Trassinelli contributed QED calculations and analysed supplementary measurements which characterize the Bragg crystal. D. S. Covita performed the frequentist analysis. L. M. Simons developed and performed the Bayesian analysis. M. Theisen contributed to the Bayesian analysis. D. S. Covita, D. F. Anagnostopoulos, H. Fuhrmann, D. Gotta, A. Gruber, A. Hirtl, T. Ishiwatari, P. Indelicato, E.-O. Le Bigot, M. Nekipelov, J. M. F. dos Santos, Ph. Schmid, L. M. Simons, M. Trassinelli, J. F. C. A. Veloso, and J. Zmeskal contributed in carrying out the experiment. D. Gotta wrote the manuscript and adapted the figures embedding contributions from V. E. Markushin, V. N. Pomerantsev, V. P. Popov, and L. M. Simons. All authors were involved in the discussion of the final manuscript.


\section*{Appendix}
\label{App:App}
\section*{A~Parameter determination} 
\subsection*{A.1~Frequentist method}\label{App:A1}

In a frequentist analysis, the extraction of information from measured data uses the so-called likelihood function $\mathcal{L}$, which describes the probability that the data are reproduced given a model with a certain set of parameters. It is to be understood as a probability density function evaluated with the data, but viewed as function of the parameters. As a distinctive feature of the frequentist method, no prior information of the parameter values is used. The evaluation of the maximum of the likelihood is supposed to determine the best values of the parameters of the model given\, \cite{Ber12}.

In the earlier evaluation\,\cite{Cov09}, this procedure was followed using the MINUIT package\,\cite{Jam75} to fit the data. It gives results for the  model parameters together with error values and a goodness of the fit estimate. 

In the present experiment, the data are binned counts ("histogram") which mostly follow a Poisson statistics. The use of a straightforward $\chi^{2} $ analysis  as resulting from a least squares analysis was claimed to be valid only for normally distributed data\,\cite{Sto97} and, therefore, had not be used for the analysis. In a basic study by Baker and Cousins \cite{Bak84} a comparison of the $\chi^{2}$ analysis with likelihood ratio tests has been performed for Poisson distributed data as well as for multinomial distributions. A so-called Poisson likelihood  $\chi^{2}$ has been proposed as best choice. In a more recent investigation by Hauschild and Jentschel\,\cite{Hau01} a maximum likelihood evaluation including likelihood ratio tests was compared to different versions of a $\chi^{2}$ analysis. As a result, it was again stated that a maximum likelihood estimator derived for Poisson statistics was optimum for the extraction of fit parameters as well as for the determination of mean values of parameters and peak areas. This is in line with the recommendations of the particle data group\,\cite{Ber12}. 
 
The use of a maximum-likelihood estimator  was criticized, however, by Bergmann and Riisager when used for complicated fit functions, short fit intervals, and less than 100 counts per channel in the data\,\cite{Ber02}. These restrictions apply to the  experimental series where the present evaluation belongs to and are causing  bias problems in parameter fitting, which lead to a different attempt for data evaluation as discussed below.

\subsection*{A.2~Bayesian method}\label{App:A2}
\subsubsection*{A.2.1 Background}\label{App:A21}

Whereas frequentist statistics provide an objective way to report the result of an experiment, the possibility to define a probability of a hypothesis or a parameter is inherent to Bayesian methods\,\cite{Siv06,Tro08}. Here, probability is considered to be conditional and is described as a degree-of-belief or plausibility, which is  associated directly  with a hypothesis or model. In consequence, a probability density function can be given for parameter values of a model after data have been taken. 

For a model (hypothesis) $M$ with a set of parameters $\mathbf{\Theta } = \{ \Theta_1 , \Theta_2 , ...\}$, the posterior probability distribution  $ Pr(\mathbf{\Theta}) \equiv Pr(\mathbf{\Theta} \mid \mathbf{D},M) $ of the model   given the data $\mathbf{D}= \{D_1 ,D_2 ,...\} $ is all what is needed  to obtain  inference about parameters $\mathbf{\Theta}$.

$Pr(\mathrm{\bf\Theta})$ is given by Bayes' theorem
\begin{eqnarray}
Pr(\mathrm{\bf \Theta}\mid {\bf D},M)=\frac{Pr({\bf D}\mid\mathrm{\bf \Theta},M)Pr(\mathrm{\bf \Theta}\mid M)}{Pr({\bf D},M)}\label{eq:Bayes_theorem}.
\end{eqnarray}

In this expression, a term for additional information $I$ in the conditional probabilities often found in literature has been omitted for the sake of transparency of the expressions. The posterior probability for the parameters $\mathbf{\Theta}$ of a model $M$ given the data is proportional to the sampling distribution of the data assuming the model (or hypothesis) being true and a prior probability function $Pr(\mathbf{\Theta}\vert M) \equiv \pi(\mathbf{\Theta})$ which encodes the state of knowledge before data have been taken. At this point, subjective belief enters data evaluation. It is therefore becoming  inevitable  to test during Bayes evaluation  whether results do depend on prior assumptions.   

In case the sampling distribution is taken as a function of the fixed ({\it i.\,e.} measured) data, it is given by the likelihood function $\mathcal{L}(\mathbf{\Theta})$ with $\mathcal{L}(\mathbf{\Theta})\equiv Pr(\mathbf{D} \vert \mathbf{\Theta},M)$. For Poisson statistics, valid in the present case, the likelihood and the logarithm of the likelihood is described by\,\cite{Siv06}
  \begin{eqnarray}
  \mathcal{L}(m_{k},d_{k})       &=& \prod_{k=1}^{N}\frac{d_{k}^{m_k}e^{-d_k}}{m_{k}!} \label{eq:L} \\
  \ln(\mathcal{L}) &=& const + \sum_{k=1}^N [m_k \ln(d_k)-d_k]\label{eq:logL}
  \end{eqnarray}
for $N$ data points $k$, $k=1, \ldots, N$, with intensity $m_k$ and the corresponding theoretical values $d_k$.

In the usual application of the Bayes method the denominator $Pr({\bf D},M)$ (see\,(\ref{eq:Bayes_theorem})) is considered  a normalization factor. For discrete parameter values it is given by the sum (law of total probability) and for continuous values by integrating the posterior over the parameter space
\begin{equation}
 \mathcal{Z} \equiv  Pr(\mathbf{D} \vert M)= \int_{\Omega_{\mathbf{\Theta}}} \mathcal{L}(\mathbf{\Theta}) \pi(\mathbf{\Theta}) d\mathbf{\Theta}.\label{eq:Z}
\end{equation}
Hence, it can be omitted in parameter estimation which can be performed  equally well from an unnormalised posterior distribution. In statistics literature, the quantity $\mathcal{Z}$ is called marginal likelihood. In cosmological applications it has been given the name of Bayesian evidence, a term which is adopted here as well. Bayesian evidence, quoted here as {\bf evidence}, can be interpreted as an average of the likelihood under the prior for a specific model choice. So it is larger in case of a model with more of its parameters being likely and smaller for a model with a more extended area in its parameters for low values of likelihood even if the likelihood function is more sharply peaked (Ockham$'$s Razor). 
 
Once the evidence given, it can be used to determine the model$'$s  posterior probability by using Bayes theorem 
\begin{equation}
 Pr(M  \vert \mathbf{D} ) \propto    Pr(\mathbf{D} \vert M) \pi(M), 
\end{equation}
where $\pi(M)$ is the prior probability assigned to the specific model. An irrelevant normalization constant has been dropped here again. 
 
Given two models $M_1$ and $M_2$, it is of interest to determine which model is supported more by the data. This question can be answered in comparing the posterior probabilities of model $M_1$ to model $M_2$ which leads to 
\begin{equation}
\frac{Pr(M_1  \vert \mathbf{D} )}{Pr(M_2  \vert \mathbf{D} )} = B_{1,2 }\cdot\frac{\pi(M_1)}{\pi(M_2)}.
\end{equation}
 $B_{1,2 }$ is called Bayes factor \cite{Kas95} and is the ratio of the evidences of the two models
\begin{equation}
B_{1,2 }  = \frac{ Pr(\mathbf{D} \vert M_1)}{ Pr(\mathbf{D} \vert M_2)}.
\end{equation}
The Bayes factor is a direct means to decide upon relative weights of models as supported by the data. It uses the ansatz that no preference for one of the models is given beforehand, {\it i.\,e.} $\frac{\pi(M_1)}{\pi(M_2)}=1$.

With the Bayesian evidences for different models being found, the Bayes factor can be used to order different models corresponding to the probability by which they are supported by the data. Such a ranking of different models avoids the need to specify different significance values as in $\chi^{2}$ tests. The difference in probability can be used to attribute significance levels to the different models following a scheme proposed by Jeffreys (Jeffreys' scale)\,\cite{Jef61,Kas95}.

One direct application of Bayesian evidence is to use them directly as weights in doing a weighted averaging over the different posterior distributions such allowing the extraction of "model free" parameters\,\cite{Mad94}.

\subsubsection*{A.2.2~Numerical approach in the Bayesian method}\label{App:A22}

Calculation of the evidence requires the solution of a many-dimensional integral. Here, the recently developed method of nested sampling was used \cite{Ski04}. This algorithm offers an efficient way to calculate the many-dimensional integral\,(\ref{eq:Z}) constituting the evidence $\mathcal{Z}$ by converting it into a one-dimensional integral 
\begin{equation}
\mathcal{Z}  =   \int_0^1 {\mathcal{L}(X)} dX.
\end{equation}
Extensive treatments are given in\,\cite{Muk06,Fer08,Fer09,Fer13a,Fer13b}. 

We follow closely the approach of Mukherjee et al.\,\cite{Muk06}, which is described roughly in the following. $X$ is the so-called prior mass with $d X = Pr(\mathbf{\Theta} \vert M ) d\mathbf{\Theta}$ and $\mathcal{L}$ is the likelihood $\mathcal{L} = Pr(\mathbf{D} \vert \mathbf{\Theta} , M)$.  The algorithm samples the prior a large number of times (typically several hundred times) with an equal prior mass given to each sample. The samples are ordered according to their likelihood and the evidence is obtained as a sum of the sequence 
\begin{equation}
 \mathcal{Z}= \sum_{j=1}^m \mathcal{L}_j w_j\label{eq:Zsum}
\end{equation}
with 
\begin{equation}
w_j =  (X_{j-1} - X_{j+1} )/2\,.
\end{equation}
The nested sampling algorithm performs the summation of (\ref{eq:Zsum}) by using a sampling/replenishment cycle. At each step of the cycle the sample with the "smallest" likelihood $\mathcal{L}_0$ is removed (and stored) and replaced by one with likelihood $\mathcal{L}>\mathcal{L}_0$. Such an algorithm seeks its way up to the region of highest likelihood by itself.

This process is stopped at a user-defined value for the instantaneous evidence which can be surveyed during the evaluation process. The still to be expected value of the evidence can be estimated and the remaining  sampling points are being stored together with the discarded points. The replacement procedure is rather delicate and has lead to different approaches. A method to treat complicated likelihood functions with several maxima is available as a program package\,\cite{Fer08,Fer09}. The present work has used the ideas behind this.

Inferences from the posterior distribution are feasible by using the total of remaining sample points plus discarded points. Each such point is assigned a weight  
\begin{equation}
p_i =  \frac{\mathcal{L}_i w_i}{\mathcal{Z}}, 
\end{equation}
from which sample-based estimates of required posterior parameters like means, credibility intervals (standard deviations) etc. can be obtained\,\cite{Siv06}. 


\section*{B~Deconvolution}
\label{App:B}

The deconvolution of the Doppler broadening from the measured line shape of the K$\beta$ transition in the $\mu$p atom presents a well-known mathematical problem and is treated as inverse problem in the theory of integral equations. It is discussed in \cite{NumRec} implying the use of a priori information. Finding the solution involves the trade-off between two optimizations: the agreement between the data and the solution on one hand and the stability or smoothness of the solution on the other hand. 

According to \cite{NumRec} this approach has a natural Bayesian interpretation: The optimization of the trade-off is equivalent to the maximization of a posterior probability function which is a product of a likelihood function and an a priori probability function. The  likelihood function describes the agreement between the data and the theoretical line shape. The data are assumed to be Poisson distributed.

\subsection*{B.1~Distributive law and maximum-entropy approach}\label{App:B1}

\begin{figure}[t]
\begin{center}
\hspace*{-2mm}
\resizebox{0.35\textwidth}{!}
{\includegraphics{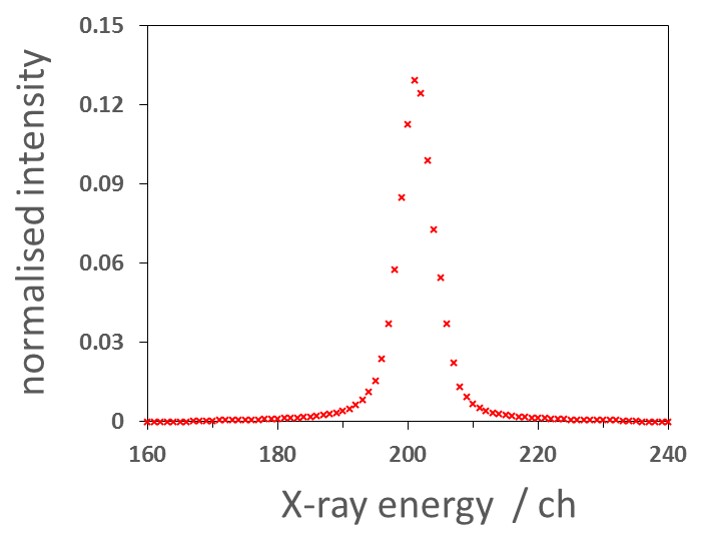}}
\caption{"Folded histogram" for a test spectrum used for deconvolution.}
\label{fig:dec1} 
\end{center}
\end{figure}

The following makes use of the fact that the resolution function has been determined with a sufficiently high precision\,\cite{Ana05,Tra07,Covth}. This permits to successfully exploit  the distributive law valid for the folding process: the folding of a histogram with a resolution function is equal to the sum of the folding of each of the bins of the histogram  with the resolution function. Formally, the theoretical values $d_{k}$ (see (\ref{eq:logL})) are then given by
  \begin{equation}
    d_k=\sum_{t=1}^N r_{k,t } a_t+ b_k, 
   \end{equation}
where $r_{k,t}$ represents a "folded bin" (normalised to one) resulting from a folding of a rectangular function characterized by the bin width of the spectrum with the resolution function representing the crystal spectrometer imaging. The values $a_t$ are the fitting parameters and the $b_k$ represent the background per bin which is assumed to be constant. In order to obtain a reliable fit value for the background, the range of the value for $t$ should largely overlap the range of the values for $k$, thus determining the background outside the distribution to be unfolded. The values $r_{k,t}$ are shown in fig.\,\ref{fig:dec1} for an arbitrary value of $t=200$ implying the hyperfine parameters of the K$\beta$ transition. It is obtained for different centres of weight $t$ by shifting this histogram to the corresponding value. In consequence, the theoretical expression $d_k$ here is, apart from the value for the background, a sum over folded histograms which are provided first and only once at the beginning of the evaluation. 

A maximum entropy prior seems appropriate as a prior probability factor for the present problem, because the kinetic energy function consists of a narrow sharp low-energy peak plus a smooth extended structure. This leads to an expression for maximum entropy prior (see equation 6.12 in\,\cite{Siv06})
  \begin{equation}
    prob(\lbrace a_t\rbrace \vert  \lbrace m_t \rbrace , \alpha, I) \propto  e^{\alpha S}
  \end{equation}
Here $m_t$ stands for a default model, $I$ for any  background information, $\alpha$ is a nuisance parameter, and $S$ is a modified Shannon-Jaynes expression as described in \cite{Siv06}  (formula 6.19) which is
  \begin{equation}
  S= - \sum_{t=1}^N (a_t \log\frac{a_t}{m_t} + (1-a_t) \log \frac{1-a_t}{1-m_t}). 
  \end{equation}

This form is specially suited for the present problem as
it allows to quantify two requests typical for a Doppler spectrum, namely that the contents of adjacent pixels decrease going outward from a central peak and that they are symmetric to it. The values for the $m_t$ are guessed initially and are updated in an iterative process. The value for the background was varied assuming a Gaussian distribution with a $\sigma$ value reflecting the possible range of error in the background determination. The value for $\alpha$ was varied between values of 0.1 to 200. 

The maximum for the posterior probability function is searched by a Monte-Carlo routine for pre-chosen values of the $a_t$, where the values for the central peak are chosen in wide limits and the values for the side channels in limits as described above. The $m_t$ values are initially assumed to be about $50\%-80\%$ of the values for $a_t$. The wide limits for the different parameters in the prior distribution guarantee a final result independent from the original assumptions.

Of the order of thousands of samples $\{a_t\}$ are accumulated together with the corresponding values for the background  parameter. They serve as input for a nested sampling routine which calculates the evidence as defined in app.\,A.2.1 at each step of the evaluation. A stopping criterion is most easily applied along the ideas presented in \cite{Muk06,Fer08,Fer09,Fer13a,Fer13b} as the terms in the evidence evaluation go exponentially to zero. A change in the value for the evidence of less than 1\% was used here. With the values for the evidence estimates, posterior quantities are easily obtained. This resulted in the mean values for $a_t$ and the background $b$ as well as in values for the corresponding errors.

\subsection*{B.2:~Deconvolution of simulated spectra}
\label{App:B2}

\begin{figure}[t!]
\begin{center}
\hspace*{-2mm}
\resizebox{0.35\textwidth}{!}
{\includegraphics{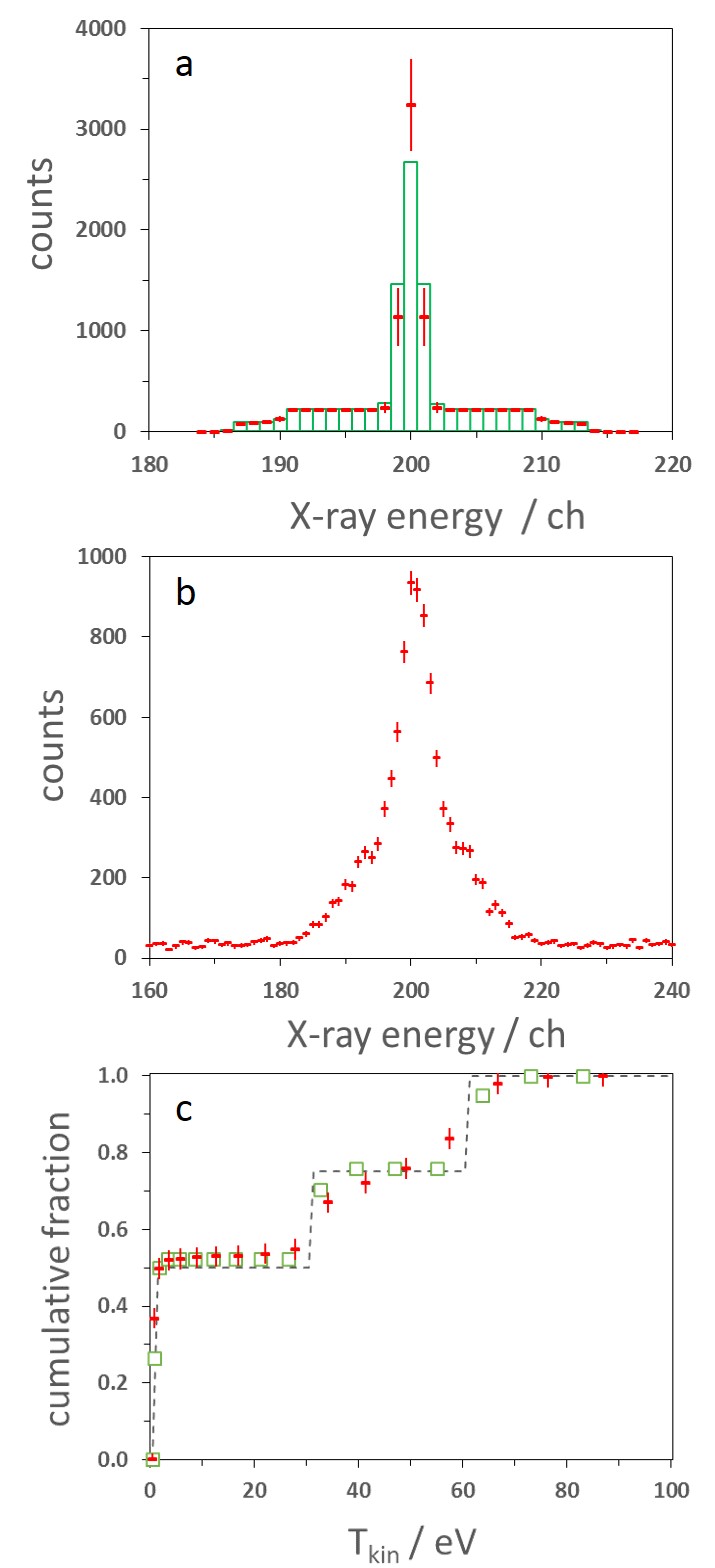}}
\caption{Test of deconvolution with a spectrum corresponding to a three-component kinetic energy distribution. a -- Doppler spectrum (histogram) and result of the deconvolution (points with error bars). One channel is equivalent to three CCD pixels. b --  Simulated X-ray spectrum derived from the Doppler spectrum with experiment set-up, intensity and background equivalent to the measured spectrum. The range is extended to the left and right to display the background conditions. c -- Cumulated energy distributions as obtained from the de-convoluted Doppler spectrum (points with error bars), as derived from the input Doppler spectrum (squares), and for the three-component test model (dashed line). All error bars represent the 1$\sigma$ level.}
\label{fig:dec4} 
\end{center}
\end{figure}

In order to give a simple example, the method was applied to a single line for a model distribution of kinetic energies with 50\% of the atoms having energies between 0 and 1\,eV and 25\% each between $30-31$\,eV and $60-61$\,eV, respectively. For the X-ray energy in question ($\mu^{-}p$ K$\beta$) this leads to a Doppler spectrum (fig.\,\ref{fig:dec4} -- a) that was adapted to the parameters of the experimental set-up by means of an X-ray tracking program (see sec.\,\ref{subsec:resolution}) yielding a spectrum equivalent to the experimental information (fig.\,\ref{fig:dec4} -- b).

Each channel corresponds to three pixels of the CCD detector, {\it i.\,e.} to 120\,$\mu$m, which means a re-binning of the measured spectrum by a factor of three. This re-binning is  result of a trade-off between noise  reduction and efficiency in the determination of the evidence on one hand and the loss of information by a higher binning factor on the other hand. 

The result of the deconvolution, which has the same binning as the simulated spectrum is shown in fig.\ref{fig:dec4} -- a (points with error bars). The comparison with the input Doppler spectrum (histogram) shows a reasonable agreement, however,  with a difficulty which is most pronounced for the central peak. This can be attributed to the drastic change in intensity for the central peak when compared to the neighbouring channels together with the rather gross bin structure. 

\begin{figure}[t!]
\begin{center}
\hspace*{-2mm}
\resizebox{0.36\textwidth}{!}
{\includegraphics{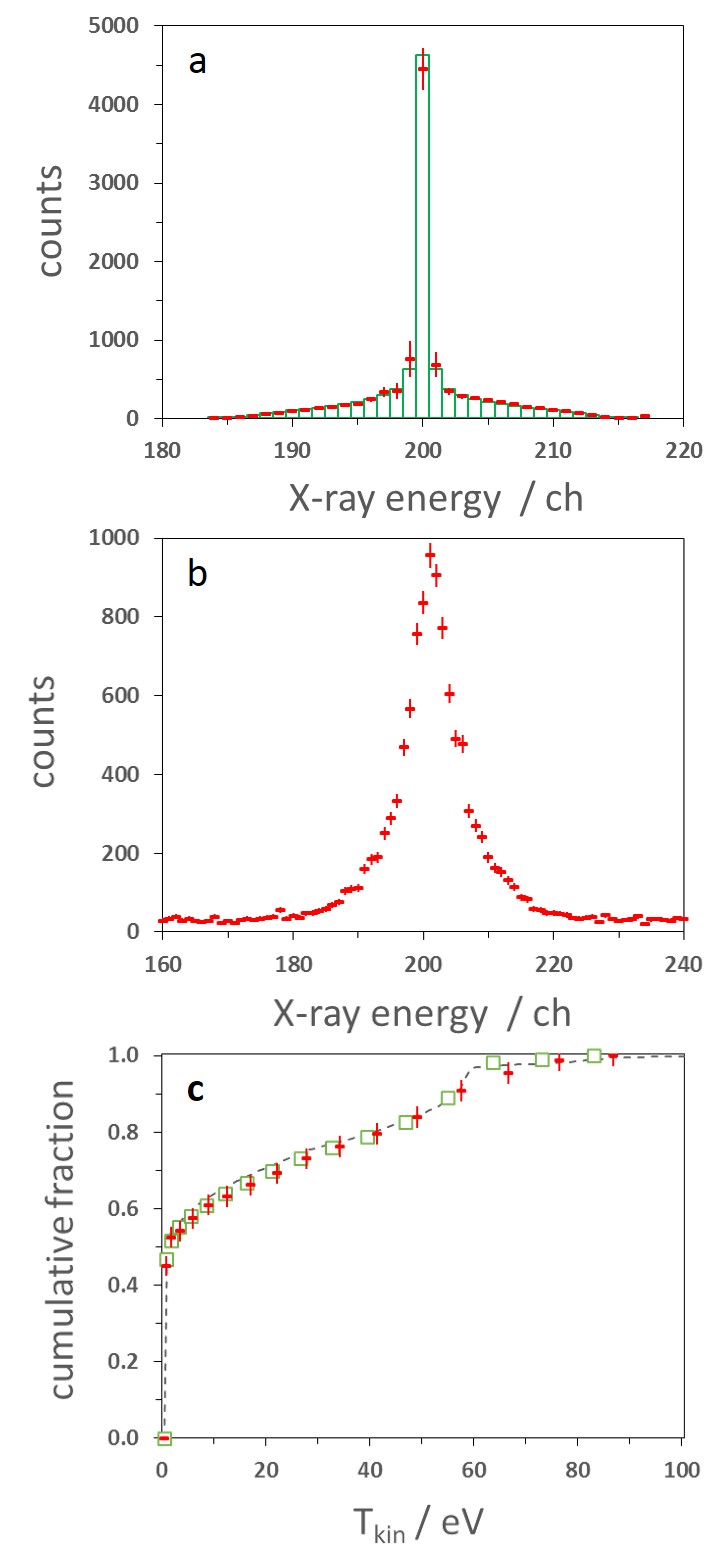}}
\caption{Deconvolution of a spectrum based on the prediction from cascade theory\,\cite{Pop11}. a -- Result of the deconvolution (points with error bars) shown together with the simulated Doppler spectrum (histogram). b -- Simulated X-ray spectrum corresponding to the experimental conditions. c -- Cumulated energy distributions obtained from the deconvolution (points) compared to the ones derived from the input Doppler spectrum (squares) and from cascade theory (dashed line).}
\label{fig:dec2} 
\end{center}
\end{figure}

In order to avoid normalization problems when going to kinetic energy distributions, a cumulative distribution was calculated as used in \cite{Bad01} and proposed for this experiment in \cite{PSI98}. It is shown in fig.\,\ref{fig:dec4} -- c and displays the steps of the three components at about the right  energies. Moreover, it correctly reproduces the intensities at  the three different energies. It shows, however, again the difficulty to cope with the sharp edges, which becomes obvious by comparing the cumulative energy distributions derived from the Doppler spectrum with the one originating directly from the kinetic energy distribution. Note the non-equidistant binning of the cumulative energy distributions, which stems from the non-linear transformation of the velocity to the kinetic energy.

The difference of the two Doppler spectra (fig.\,\ref{fig:dec4} -- a) may be quantified by means of a $\chi^2$ analysis. The comparison of the distributions obtained from the deconvolution of simulated data and the theoretical Doppler spectrum as input for the simulation yields a reduced 
$\chi_{\mathrm{red}}^2=0.87$ with a number of degrees of freedom $n.d.f.=32$.

To apply the deconvolution method to a spectrum as close as possible to the measured spectrum, the theoretical kinetic energy distribution given by\,\cite{Pop11} was used.  In addition, the theoretical hyperfine structure of the K$\beta$ transition is taken into account for this Doppler spectrum  (fig.\,\ref{fig:dec2} -- a (histogram)). The result of the tracking routine for the X-ray line of 2249.6\,eV simulating the experimental conditions is displayed in fig.\,\ref{fig:dec2} -- b.  

The result of the de-convolution is shown in fig.\,\ref{fig:dec2} -- a (points with error bars). The cumulative energy distributions corresponding to the theoretical Doppler spectrum (squares), to the deconvolution of the data (points), and directly from the cascade calculation (dashed lines) are shown in fig.\,\ref{fig:dec2} -- c. A good agreement can be stated. The $\chi^2$ analysis of the two Doppler spectra (fig.\,\ref{fig:dec2} -- a) yields $\chi_{\mathrm{red}}^2=0.91$ with $n.d.f=32$.


\end{document}